\documentclass[fleqn,usenatbib]{mnras}

\usepackage{newtxtext,newtxmath}
\usepackage[T1]{fontenc}

\DeclareRobustCommand{\VAN}[3]{#2}
\let\VANthebibliography\thebibliography
\def\thebibliography{\DeclareRobustCommand{\VAN}[3]{##3}\VANthebibliography}

\usepackage{xcolor}
\usepackage{hyperref}
\newcommand{\corrauthmark}{%
  \protect\hyperlink{corr-author-note}{\footnotemark[1]}%
}

\usepackage{textgreek}
\usepackage{graphicx}
\usepackage{amsmath}
\usepackage{siunitx}
\usepackage{multirow}
\usepackage{lineno}

\usepackage{cleveref}
\crefname{section}{Section}{Sections}
\crefname{figure}{Figure}{Figures}
\crefname{table}{Table}{Tables}
\crefname{equation}{Equation}{Equations}

\usepackage{threeparttable}
\usepackage{natbib}

\newcommand{\kms}{~km~s$^{-1}$}
\newcommand{\ps}{\,{\rm s}^{-1}}
\newcommand{\twCO}{$^{12}$CO}   
\newcommand{\thCO}{$^{13}$CO}
\newcommand{\Jotz}{$J$=1--0}

    \newcommand{\km}{\,{\rm km}}

\title{1LHAASO J1852+0050u: GeV-100TeV Gamma-ray emission Powered by Star-forming Region? }

\author[LHAASO Collaboration]{
Zhen Cao,$^{1,2,3}$
F. Aharonian,$^{3,4,5,6}$
Y.X. Bai,$^{1,3}$
Y.W. Bao,$^{7}$
D. Bastieri,$^{8}$
X.J. Bi,$^{1,2,3}$
Y.J. Bi,$^{1,3}$
\newauthor
W. Bian,$^{7}$
J. Blunier,$^{9}$
A.V. Bukevich,$^{10}$
C.M. Cai,$^{11}$
% ...
\\  (The LHAASO Collaboration) 
Affiliations are listed at the end of the paper 
}

\date{Accepted XXX. Received YYY; in original form ZZZ}

\pubyear{\the\year{}}

\begin{document}
\label{firstpage}

\pagerange{\pageref{firstpage}--\pageref{lastpage}}
\maketitle


\begin{abstract}
Star-forming regions (SFRs), including HII regions, have recently attracted significant attention as potential sites of particle acceleration capable of producing PeV cosmic rays, i.e.\ as potential Galactic PeVatrons.
Yet, SFRs known as candidate PeVatrons are still very rare.
Here we 
report a highly likely physical association between the UHE gamma-ray source 1LHAASO J1852+0050u and coincident star-forming regions. 
Using $\sim$4 years of LHAASO data, the extended source 1LHAASO J1852+0050u (called J1852 in this study) was reanalyzed with a significance of $\sim13\sigma$ and 39\%-containment radius of $0.35\degr$. This extended source projectively covers dozens of H\textsc{ii} regions and an energetic pulsar PSR~J1853+0056.
We reanalyzed $\sim16$ years of \textit{Fermi}-LAT data and found a GeV source (named SrcA$\rm _G$) coincident with the LHAASO source J1852.
The CO-line observation shows a molecular clump which coincides with the gamma-ray sources and is related with the ultra-compact H\textsc{ii} region G34.26+0.15 at the distance of $\sim3.3$~kpc.
Fitting to the spectral energy distribution indicates that PSR~J1853+0056 cannot afford the observed gamma-ray flux either for J1852 or SrcA$\rm _G$.
Although the contribution from hidden pulsars to the gamma-ray emission could not be excluded, it is most likely that the GeV--100TeV gamma-ray emission is dominantly powered by SFR through p-p hadronic interaction between the protons accelerated in the SFR and MCs: the GeV emission is
ascribed to the protons accelerated by protostars' activities, while 
the TeV emission 
ascribed to the protons accelerated by massive stars 
in the SFR.
\end{abstract}

\maketitle

\section{Introduction}      \label{sec:intro}

The origin of Galactic cosmic rays (CRs), particularly those reaching energies up to the knee at a few PeV, remains one of the most fundamental open questions in high-energy astrophysics.
Current understanding suggests that the CRs with energies below the ``knee" ($3\times10^{15}$~eV) are primarily produced within the Milky Way \citep{1934PNAS...20..259B,1964ocr..book.....G}.
The idea that some Galactic sources can accelerate particles up to PeV energies has 
stimulated extensive gamma-ray studies on supernova remnants (SNRs), pulsars, young massive star clusters, star-forming regions (including H\textsc{ii} regions), etc.
SNRs have long been regarded as the primary accelerators of Galactic CRs through diffusive shock acceleration \citep[e.g.,][]{1978MNRAS.182..147B}. However, accumulating observational evidence suggests that SNRs alone may be insufficient to account for the highest-energy Galactic CRs, motivating the exploration of additional or complementary acceleration sites within the Milky Way.


Recently, SFRs, including HII regions, have attracted more and more attention as potential sites of particle acceleration capable of producing UHE gamma rays.
Given the frequent and violent stellar activity of O- and B-type stars in their interiors, SFRs have been suggested to be capable of accelerating high-energy particles via colliding winds in massive binaries \citep{2020A&A...635A.167H}, as well as via wind-driven shocks and turbulence produced by the collective stellar winds of young massive clusters
\citep[e.g.,][]{2019NatAs...3..561A,2022ChPhC..46c0002C,2025icrc.confE1390P}.
In addition, young stellar objects (YSOs), including protostellar sources, have also been proposed as sites of particle acceleration to relativistic energies associated with their jets \citep[e.g.,][]{2016A&A...590A...8P,2023MNRAS.523..105D,2022RAA....22b5016Y}.

Observational evidence, especially from gamma-ray band, increasingly supports the role of SFRs as CR accelerators.
Extended gamma-ray emission associated with massive star clusters or H\textsc{ii} regions has not only been reported at GeV energy range \citep[e.g.,][]{2011Sci...334.1103A,Yang2018Westerlund2,2022A&A...659A.101L,2024MNRAS.535.1526L}, but also found in TeV energy range  \citep[e.g.,][]{Aharonian2007Westerlund2,Aharonian2022Westerlund1,2021NatAs...5..465A}.   
At even higher energies, several prominent SFRs and H\textsc{ii} regions, such as Cygnus OB2 \citep{2021NatAs...5..465A}, G35.6$-$0.5 \citep{2025ApJ...979...70C}, and W43 \citep{2025SCPMA..6879502C}, have been detected by LHAASO at energies approaching or exceeding 100 TeV, indicating that star-forming environments may host PeVatron candidates.


%

As listed in the first LHAASO source catalog, 1LHAASO J1852+0050u was detected above 100~TeV with a test statistics (TS) value of TS$_{100}\geq$20 \citep{2024ApJS..271...25C}.
With 39\%-containment $r_{39}=0.64\pm0.07\degr$ for WCDA and $r_{39}=0.85\pm0.06\degr$ for KM2A, it covers two SNRs (W44 and Kes~79), multiple pulsars, and numerous HII regions.
Based on these spatial coincidences, the catalog proposed possible associations of 1LHAASO J1852+0050u with a SNR/PWN related to Kes~79 or a TeV halo powered by the middle-aged pulsar PSR~J1853+0056 \citep{2024ApJS..271...25C}.
In this work, we examine both scenarios and find that neither can provide a dominant explanation for the observed gamma-ray emission.
The large number of HII regions within 1LHAASO J1852+0050u region therefore motivates us to explore star-forming activity as an alternative origin of the gamma-ray emission.
Among these H\textsc{ii} regions, the G34.26+0.15, which is a massive star-forming region and consists of four radio components (A, B, C, and D) \citep{1985ApJ...288L..17R}, is very bright in infrared and radio bands.
The local-standard-of-rest (LSR) velocities of G34.26+0.15 was fitted to range from +53 to +58\kms using the CS(2-1) line \citep{1996A&AS..115...81B}, H110$\alpha$ radio recombination line and H$_2$CO line \citep{2002ApJS..138...63A}, and hydrogen radio recombination line \citep{2011ApJ...738...27B}, etc.
Due to the uncertainty range of the LSR velocities for different line emissions, there is a small scattering range in the kinematic distance from 3.3 to 3.7 kpc for G34.26+0.15.
Explosive outflows were suggested in G34.26+0.15C \citep{2025AJ....169..324I}, a cometary ultra-compact H\textsc{ii} region 
\citep{1985ApJ...288L..17R},
which was suggested to be a single star moving supersonically through molecular clouds (MCs) \citep{1991ApJ...369..395M,1992ApJ...394..534V}.
With the accumulation of exposure time, the contribution of different objects can be separated from the large extended source 1LHAASO J1852+0050u.

In this work, we present multi-wavelength studies towards the region of 1LHAASO J1852+0050u. The information of the observational data is described in \cref{sec:Observations}.
The results are given in \cref{sec:results}, the origin of gamma-ray emission is discussed in \cref{sec:dis}.
Finally, we conclude this study in \cref{sec:concl}.

\section{Observations and data} 
\label{sec:Observations}

\subsection{LHAASO Observational data}
\label{sec:lhaaso data}

The LHAASO data used here were collected with the full array of WCDA from March 5, 2021 to July 31, 2025 (with a livetime of $\sim$ 1484 days) and the full array of KM2A from July 20, 2021, to July 31, 2025 (with a livetime of $\sim$ 1438 days).
For WCDA, the number of triggered detectors ($N_{\rm hit}$) is divided into 6 bins: 60-100, 100–200, 200–300, 300–500, 500–800, and 800–2000. The event bins can serve as a proxy for energy binning. Utilizing the $\cal P$ parameter (termed the PINCness of an event) in \cite{2017ApJ...843...39A}, the gamma-rays from the cosmic ray background are separated.
Events with ${\cal P}<1.1$ are retained in this analysis.
The KM2A data was divided into 5 equally logarithmic bins per decade from 25 TeV to several PeV with bin size $\Delta \log_{10}E_\gamma=0.2$.
The residual cosmic ray background is estimated using the `direct integration method' \citep{2004ApJ...603..355F} for both WCDA and KM2A data.

For both WCDA and KM2A data, we chose a $6\degr\times10\degr$ square regions of interest (ROIs) centered at $l=34.5\degr$, $b=0\degr$ in the galactic coordinate system. During analysis, the sky within ROI is binned into cells of a size of $0.1\degr\times0.1\degr$ and each cell was filled with events according to their reconstructed arrival directions. Due to the existence of the Galactic diffuse emission (GDE), the dust column density measured by the PLANCK satellite \citep{2014A&A...571A..11P,2016A&A...596A.109P} was adopted to model the background gamma-rays arising from the collision between the Galactic cosmic rays and the interstellar medium.
In addition, a 3D likelihood fitting process was used to determine the spectrum and morphology simultaneously.
During the fitting process, the normalization factor and spectral index of the GDE component were set as free parameters.

\subsection{\textit{Fermi}-LAT Observational data}
\label{sec:fermi data}

We analyzed more than 16 years (from 2008-08-04 15:43:36 (UTC) to 2024-09-01 23:05:01 (UTC)) of \textit{Fermi}-LAT Pass 8 SOURCE class (evclass=128, evtype=48) data with the software Fermitools 2.2.0\footnote{\url{https://fermi.gsfc.nasa.gov/ssc/data/analysis/software/}}. The ROI in our study are $15^{\circ}\times15^{\circ}$ in size, centered at the position of PSR~J1853+0056 
(${\rm R.A._{J2000}}=283.38^{\circ}$, ${\rm Dec_{J2000}}=0.95^{\circ}$).

Firstly, the data selection was performed with command \textit{gtselect} with the maximum zenith angle of 90$^{\circ}$ to reduce the contamination from the Earth limb. Then, we applied command \textit{gtmktime} to the data with recommended filter string “(DATA\_QUAL \textgreater 0)\&\&(LAT\_CONFIG == 1)” for choosing good time intervals. The entire energy range from 0.2~GeV to 500~GeV was divided into 10 logarithmic bins per decade for counts cube and exposure cube. The appropriate Instrument Response Functions are “P8R3\_SOURCE\_V3”. The Galactic interstellar diffuse background emission model ``\textit{gll\_iem\_v07}” and isotropic background spectral template ``\textit{iso\_P8R3\_SOURCE\_V3\_v1}”, as well as the sources listed in \textit{Fermi}-LAT 14-year source catalog \cite[4FGL-DR4, ][]{2023arXiv230712546B} within a radius of $25^{\circ}$ from the ROI center, were incorporated for analysis using the user-contributed tool \textit{make4FGLxml.py}\footnote{\url{https://fermi.gsfc.nasa.gov/ssc/data/analysis/user/make4FGLxml.py}}. 
After that we used the python module \textit{pyLikelihood} with the NEWMINUIT optimizer to perform the binned likelihood analysis and got the best-fit results. In this step, we only freed spectral parameters of the catalog sources within 5$^{\circ}$ from the ROI centers and the normalization of the two diffuse background components.
In addition, a python package Fermipy \citep{2017ICRC...35..824W} was employed
\footnote{\url{https://fermipy.readthedocs.io/en/latest/}} (version 1.3.1) in the position, extension, and spectral energy distribution (SED) fitting process, during which all event types (evtype=3) were adopted.

\subsection{Molecular Data}

The archival data of the \twCO\ \,(\Jotz) line at \SI{115.271}{GHz} and the \thCO\ \,(\Jotz) line at \SI{110.201}{GHz} of the FOREST Unbiased Galactic plane Imaging survey with the Nobeyama 45 m telescope \cite[FUGIN, ][]{FUGIN}) observation were used to analyze the molecular environment.
The angular resolution were 20$^{\prime\prime}$ and 21$^{\prime\prime}$ for \twCO\ and \thCO.
The average rms noise were $\sim$\SI{1.5}{\kelvin} and $\sim$\SI{0.7}{\kelvin} for \twCO\ and \thCO\ at a velocity resolution of 0.65\,km\,s$^{-1}$, respectively.

\subsection{Other data}

We also used the 1.3~GHz radio continuum data from SARAO MeerKAT Galactic Plane Survey \cite[SMGPS, ][]{2024MNRAS.531..649G}. The radio continuum image has a spatial resolution of $\sim8^{\prime\prime}$ and a sensitivity of $\sim10$--20 $\mu$Jy/beam.

\section{Data analysis and results} \label{sec:results}
\subsection{LHAASO Data Analysis} \label{sec:lhaaso}

\begin{figure*}
  \centering
	\includegraphics[width=8.2cm]{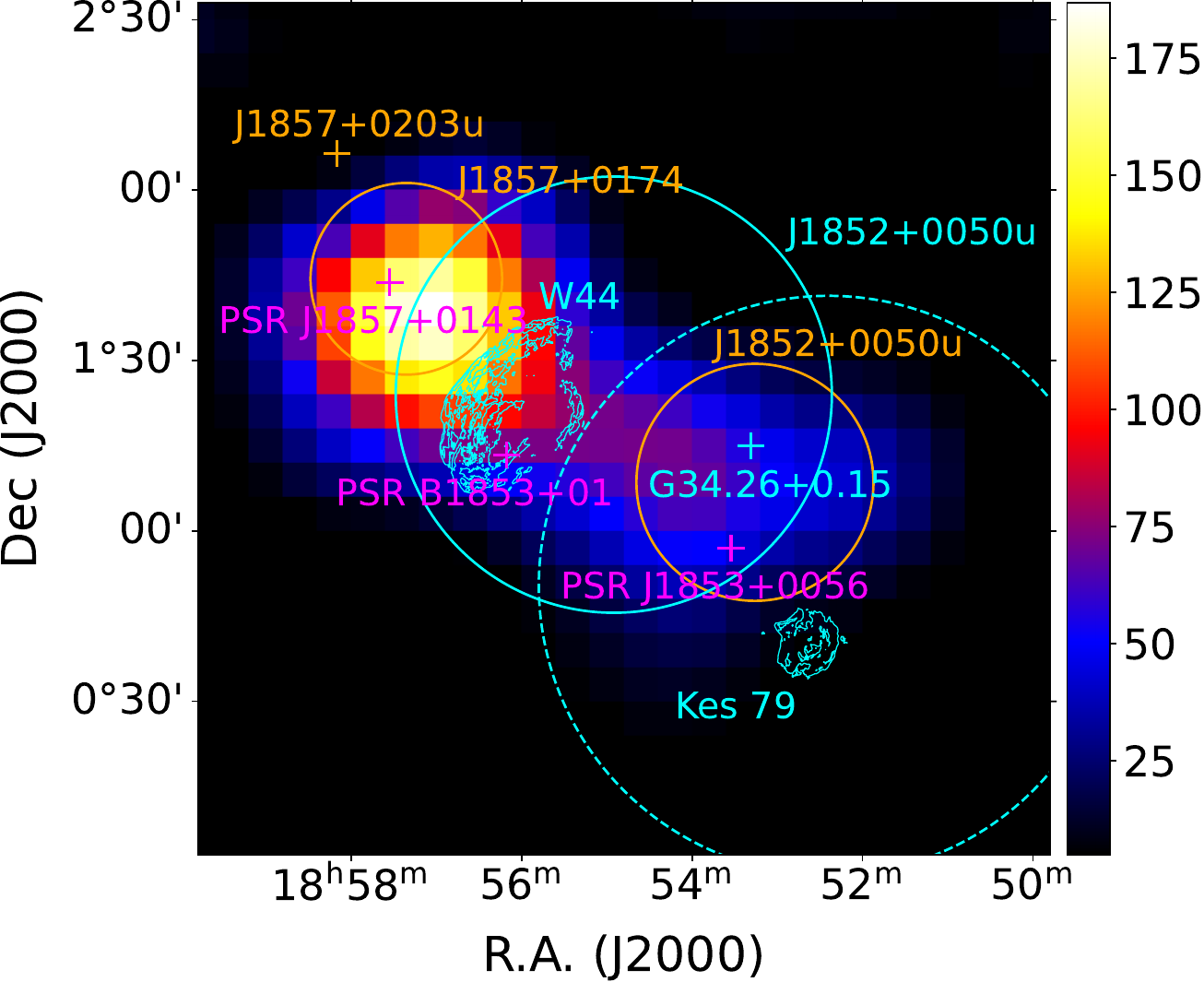}
    \includegraphics[width=8cm]{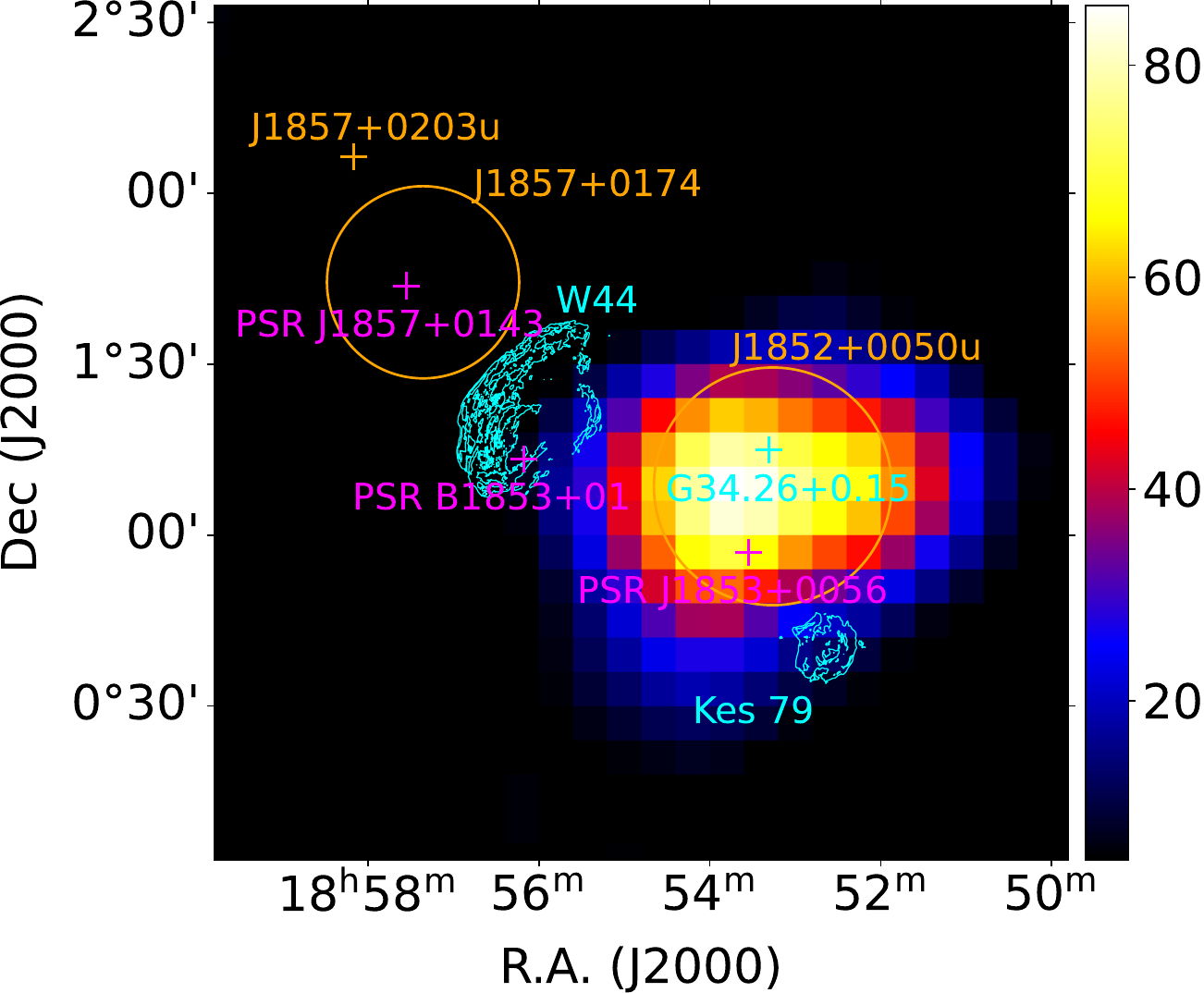}
  \caption{TS maps of 2.5$^{\circ}\times2.5^{\circ}$ regions above 1 TeV towards the 1LHAASO J1852+0050u.
  Left panel: the TS map by refitting the LHAASO catalog source \protect\cite{2024ApJS..271...25C} but with the source 1LHAASO J1852+0050u removed from the model.
  The position and extension of 1LHAASO J1852+0050u in the LHAASO first catalog were shown in cyan solid (WCDA) and dashed (KM2A) circles for comparison.
  Right panel: the TS map of J1852 with the contribution from LHAASO J1857+0174 subtracted.
  The cyan plus indicates the star-forming region G34.26+0.15.
  SNRs W44 and Kes 79 are delineated as cyan contours with [5, 10] and [1, 6] mJy beam$^{-1}$ levels of the 1359.7 MHz radio continuum emission from MeerKAT observations  \citep{2024MNRAS.531..649G}, respectively.
  The orange circles represent the 1$\sigma$ Gaussian radius of J1852 and LHAASO J1857+0174.
  The pulsars with spin-down luminosity $\geq10^{34}$erg s$^{-1}$ were depicted in magenta pluses.}
\label{fig:tsmap_TeV}
\end{figure*}

Using the updated LHAASO data with livetime almost 3 years longer than in \cite{2024ApJS..271...25C}, we found that the source 1LHAASO J1852+0050u can be modeled naturally as an ellipse source or two Gaussian sources whichever for WCDA or KM2A.
A joint analysis which combines the data of two arrays were applied.
During the analysis, the normalization, index, position, and extension of sources in the ROI (such as 1LHAASO J1857+0203u) were freed.
We first generated a test-statistic (TS) map by excluding 1LHAASO J1852+0050u from the catalog model in the energy range of $\geq1$ TeV (see the left panel of \cref{fig:tsmap_TeV}).
The TS value for each pixel was evaluated by ${\rm TS}=-2{\rm \ln}(\cal L_{\rm0}/L_{\rm1})$, where $\cal L_{\rm0}$ is the maximum likelihood of the null hypothesis and $\cal L_{\rm1}$ is the maximum likelihood of the test model that a putative point source was located in this pixel with a fixed index of 2.7 in the energy range of 1--25 TeV and 3.2 at energies above 25 TeV.
As can be seen, the residual emissions appear to peak in the northeast and southwest of SNR W44.
We thus adopted two hypotheses to fit the residual emissions, separately. One is one-ellipse model, and the other is two-Gaussian model, in which the two Gaussian sources named as LHAASO J1852+0050u (hereafter called J1852) and LHAASO J1857+0174 \footnote{LHAASO J1857+0174 is also called W44-NE in a separate paper by LHAASO collaboration (Cao et al. 2026, submitted) with the same position and extension. Because this source is not our target, its spectral type is not tested in this work}. 

\cref{tab: spa_para} lists the best fitted spatial and spectral parameters for the two hypotheses. 
During the fitting, the spectral type of J1852 was postulated to be LogParabola (LogP) spectrum (that is, $dN/dE=N_0(E/E_0)^{-\varGamma-\beta\mathrm{ln}(E/E_0)}$) with $E_0$ fixed to be 10 TeV.
The Akaike information criterion \cite[AIC, ][]{1974ITAC...19..716A} was adopted for each case to evaluate which model is preferred.
The AIC is defined as ${\rm AIC}=2k-2\ln {\cal L}$, where $k$ is the number of free parameters in the model and $\cal L$ is the maximum likelihood estimate.
By comparing AIC between different models, the one with minimum AIC performs better than others.
After subtracting the contribution from the sources, though no significant residual emission is detected under either hypothesis, the lower AIC means that the two-Gaussian model performs significantly better. 
As shown in \cref{fig:tsmap_TeV}, the J1852 (with significance of $\sim13\sigma$) is overlapped with the star-forming region G34.26+0.15 along the line of sight (LOS). So in this work, we will focus only on J1852 to explore possible CRs acceleration by embedded protostars and massive stars in G34.26+0.15.
First, we test different spatial morphology of J1852 by changing the Gaussian model into uniform Disk and diffuse model (from \citealt{2017Sci...358..911A}), separately.
The results of $\Delta$AIC$\le2$ between these three models indicate no significant difference and the Gaussian model is used in following analysis.

\begin{table*}
\caption{Results of the Spatial Analysis towards the 1LHAASO J1852+0050u Region.}
\centering
\label{tab: spa_para}
\begin{tabular}{lcccc}
\hline\hline
\textbf{Model}                 & \textbf{R.A., decl. ($\degr$)} & \textbf{Extension ($\degr$)} & \textbf{Spectral parameters} & \textbf{$\Delta$AIC} \\
\hline
Ellipse                  &$283.93\pm 0.05\degr$, $1.45\pm 0.03 \degr$ &($0.80\pm -0.05\degr$, $0.27\pm 0.04\degr$) & $N_0=18.40\pm2.10$, $\varGamma = 2.45\pm 0.08$, &  0\\
                      &                                         &                                 &$\beta=0.42\pm0.06$ & \\
\hline
Gaussian (J1852)              & $283.33\pm 0.04\degr$, $1.14\pm 0.04 \degr$  &$0.35\pm 0.03\degr$ & $N_0=10.60\pm0.79$, $\varGamma = 1.74\pm0.07$, & -46.5  \\
                      &                                              &                              & $E_{\rm cut}=23.63\pm1.29$ TeV& \\
Gaussian (LHAASO J1857+0174)              & $284.34\pm 0.03\degr$, $1.74\pm 0.03 \degr$  &$0.28\pm 0.02\degr$ & $N_0=13.70\pm0.77$, $\varGamma = 2.11\pm0.05$, &    \\
                      &                                              &                              & $E_{\rm cut}=27.34\pm1.53$ TeV& \\
\hline
\end{tabular}
\begin{tablenotes}
\item Note: the values of $N_0$ are given in units of $10^{-15}$ TeV$^{-1}$cm$^{-2}$s$^{-1}$. The extension for the Gaussian models refers to their respective 39\%-containment radii. The $1\sigma$ statistical uncertainties were given for morphological and spectral parameters. All parameters in this table were determined in the analysis of events with energies above $\sim$1 TeV.
\end{tablenotes}
\end{table*}

Once the spatial model was determined, the spectral characteristic of J1852 was studied by testing the other three spectral types PowerLaw (PL), ExpCutoffPowerLaw (ECPL), and BrokenPowerLaw (BPL) in separate fits.
The formulae and performance of these spectral types were detailed in \cref{tab:spec}.
By comparison with the PL, the other three types apparently perform better, which means obvious curvature.
The discrepancy of $\Delta$AIC between the spectra of ECPL and BPL is small, indicating that these two spectra is statistically indistinguishable.
In the following analysis, the spectral type of ECPL was adopted for the less degrees of freedom.

\begin{table*}
\caption{Formulae and Likelihood Test Results of Gamma-ray Spectra Type for J1852 above 1 TeV and SrcA$\rm _G$ in the energy range of 0.2--500~GeV.} 
\centering
\label{tab:spec}
\begin{tabular}{cllcc|cc}
\hline
\hline
\textbf{Name} & \textbf{Formula} & \textbf{Free parameters} & \multicolumn{2}{c}{\textbf{J1852}} & \multicolumn{2}{c}{\textbf{SrcA$\rm _G$}} \\
\cline{4-7}
    &   &   & Flux (erg cm$^{-2}$s$^{-1}$) & $\Delta$AIC & Flux (erg cm$^{-2}$s$^{-1}$) & $\Delta$AIC \\
\hline
PL   & $dN/dE=N_0(E/E_0)^{-\varGamma}$ & $N_0, \varGamma$ & 4.7$\times10^{-12}$ &0 & 6.0$\times10^{-11}$ & 0\\
ECPL & $dN/dE=N_0(E/E_0)^{-\varGamma}\mathrm{exp}(-E/E_\mathrm{cut})$   &  $N_0,\varGamma,E_\mathrm{cut}$ & 3.7$\times10^{-12}$ & -50.6 & 5.1$\times10^{-11}$ & -28.1\\
LogP & $dN/dE=N_0(E/E_0)^{-\varGamma-\beta\mathrm{ln}(E/E_0)}$   &   $N_0,\varGamma,\beta$  & 3.2$\times10^{-12}$ & -44.9 & 9.0$\times10^{-11}$ & -45.7\\
BPL  & $dN/dE=N_0$ 
$\begin{cases}
 (E/E_{\rm b})^{-\varGamma_1}, E\le E_{\rm b}\\(E/E_{\rm b})^{-\varGamma_2}, E\ge E_{\rm b}
\end{cases}$   &   $N_0, E_\mathrm{b}, \varGamma_1, \varGamma_2$ & 4.0$\times10^{-12}$ & -52.5 & 5.7$\times10^{-11}$ & -43.9\\
\hline
\end{tabular}
\end{table*}

\begin{figure*}
  \centering
	\includegraphics[width=8cm]{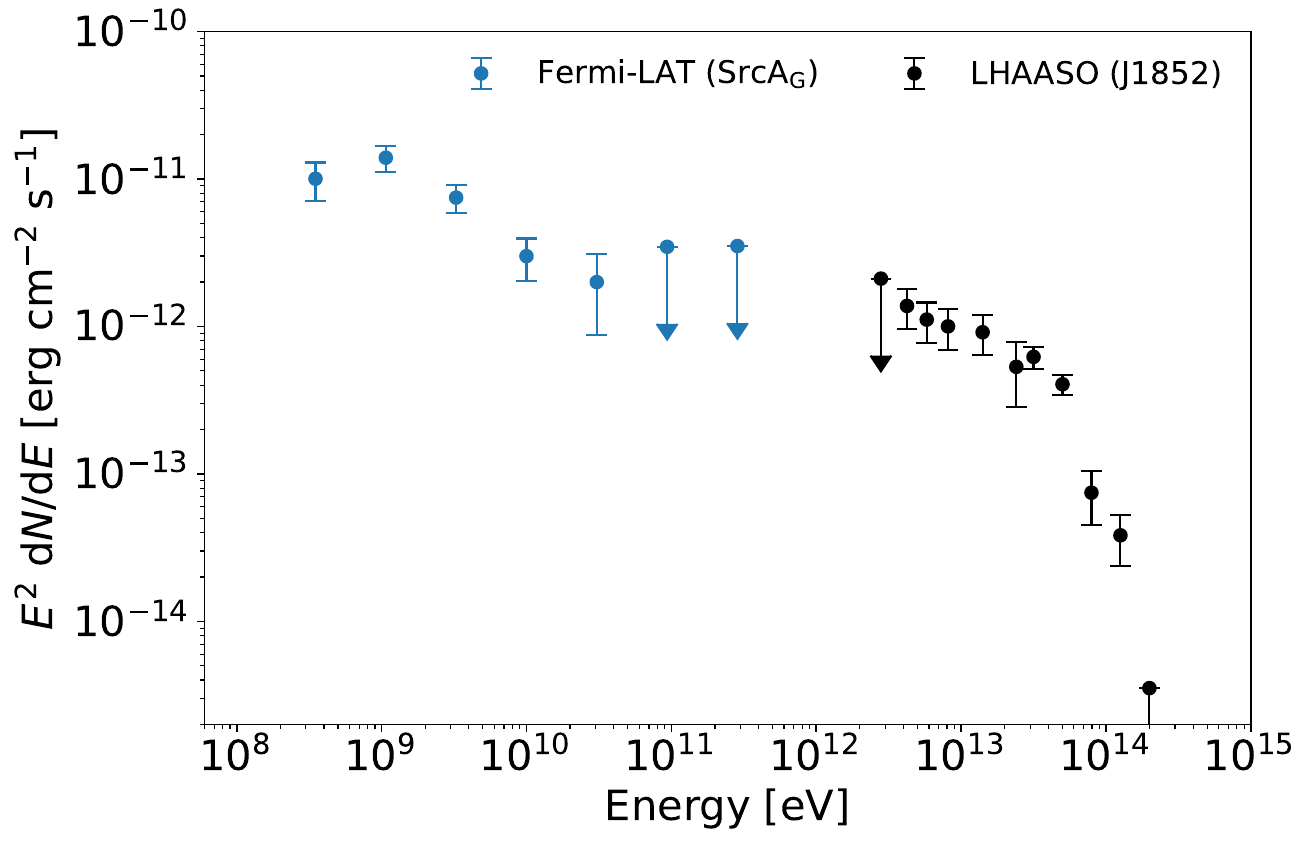}
  \caption{Spectral energy distribution of J1852 and SrcA$\rm _G$. The black and blue points represent gamma-ray flux from LHAASO and \textit{Fermi}-LAT, respectively.}
 \label{fig:sed}
\end{figure*}

The SED was constructed with the bins on the basis of the $N_{\rm hit}$ for WCDA data and $\Delta \log_{10}E_\gamma=0.2$ for KM2A data. During the fitting, we fix the spatial parameters and spectral index to the best-fit results obtained above. Only the normalization parameters are free for all the LHAASO sources within the ROI as well as the GDE. The 95\%  upper confidence limit was estimated for the energy bins with TS value less than 4.
The systematic uncertainty caused by the imperfection model of the GDE were calculated through conservatively varying the normalization of GDE by $\pm10\%$ from the best-fit values in the entire energy band.
For WCDA, the systematic uncertainty caused by energy scale are as large as 8\% on the flux \citep{2021ChPhC..45h5002A}.
For KM2A, the main systematic error is contributed by the atmospheric model in the Monte Carlo simulations and estimated to be 7\% on the flux \citep{2021ChPhC..45b5002A}.
Then the systematic errors and statistic errors were combined in quadrature.
The flux points were shown in \cref{fig:sed}.

While the 1LHAASO J1852+0050u was marked with TS$_{100}\geq$20 (the TS value for energy above 100~TeV) by \cite{2024ApJS..271...25C}, in this study, we detected a TS value of $\le10$ for J1852 above 100~TeV.
Given the large $r_{39}=0.85\pm0.06\degr$ of 1LHAASO J1852+0050u in KM2A and moderately high TS$_{100}=$22.1, the TS$_{100}$ obtained previously could be contaminated by photons from surrounding energetic sources (i.e., 1LHAASO J1857+0203u with TS$_{100}\geq$100).

\subsection{\textit{Fermi}-LAT Data Analysis} \label{sec:fermi}

Generally, the nature of gamma-ray sources can hardly be clarified using only TeV data. 
Hence, we searched for the potential counterpart of J1852 in GeV energy band with the \textit{Fermi}-LAT data to advance the understanding of this source.
During the morphological analysis, the events of PSF0 and PSF1 were discarded due to the poor quality of reconstructed direction.

\begin{figure*}
  \centering
	\includegraphics[width=16cm]{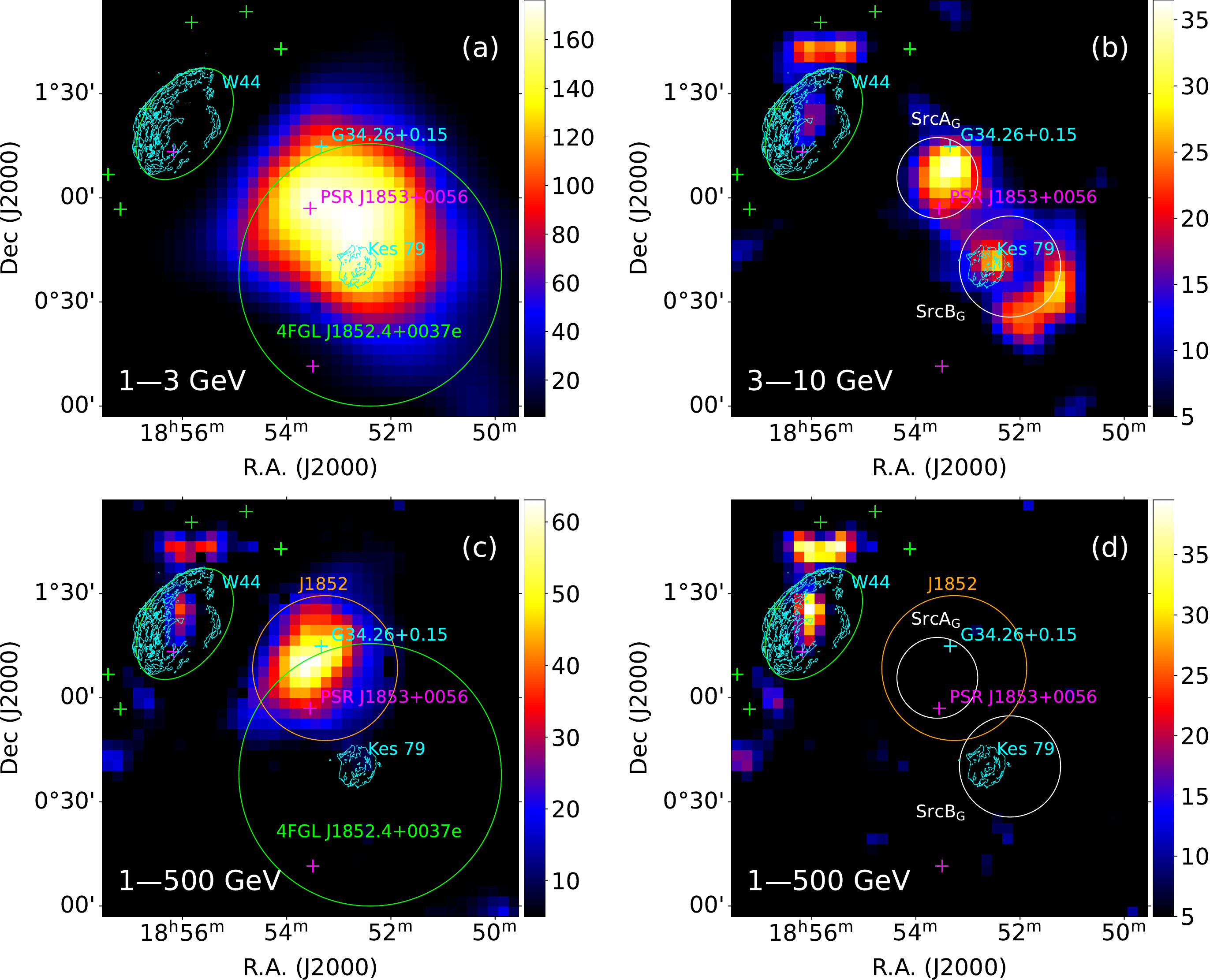}
  \caption{Fermi-LAT gamma-ray TS maps of 2$^{\circ}$ × 2$^{\circ}$ regions centered at PSR J1853+0056. (a): the TS map in the energy range 1--3~GeV for an improved angular resolution with source 4FGL~J1852.4+0037e excluded from the 4FGL-DR4 catalog sources model. (b): same as (a) but in the energy range 3--10~GeV. (c): the TS map with 4FGL-DR4 catalog sources model in the energy range 1--500~GeV. (d): the TS map using the two extended source model.
  The green pluses, circle, and ellipse mark the positions of 4FGL-DR4 catalog sources \citep{{2023arXiv230712546B}}.
  The cyan plus and contours indicate the same objects as in \cref{fig:tsmap_TeV}.
  The white circles represent 39\% extension of the best-fitted spatial models in GeV energy band.
  The orange circle represents 1$\sigma$ Gaussian radius of J1852.
  The pulsars with spin-down luminosity $\geq10^{34}$ erg s$^{-1}$ are depicted in magenta pluses.
  }
  \label{fig:tsmap_GeV}
\end{figure*}

The TS maps around the J1852 region in the energy range 1--3 and 3--10~GeV with 4FGL~J1852.4+0037e excluded from the 4FGL-DR4 catalog sources model were shown in \cref{fig:tsmap_GeV}(a) and (b).
It can be seen that the gamma-ray emissions in 3--10~GeV spatially peak close to the massive star-forming region G34.26+0.15 and the SNR Kes~79, which hints hybrid contribution of 4FGL~J1852.4+0037e from different objects.

Then, we generated a TS map in the energy range 1--500 GeV by including all of the 4FGL-DR4 catalog sources as models (see \cref{fig:tsmap_GeV}(c)).
The 4FGL-DR4 catalog placed the extended source 4FGL~J1852.4+0037e (with extension $\sim0.63\degr$), which partially overlapped with J1852, at ${\rm R.A._{J2000}}=283.10^{\circ}$, ${\rm Dec_{J2000}}=0.63^{\circ}$ (shown as the green circle in \cref{fig:tsmap_GeV}(c)).
As can be seen, some residual emission appears to the northeast of 4FGL~J1852.4+0037e, indicating that the residual emission cannot be well fitted by a single Disk model.
This extended GeV source (4FGL~J1852.4+0037e) was then divided into two uniform Disk models (Src-N and Src-S) by \cite{2022ApJ...928...89H} in terms of different energy ranges using 11.5 years of \textit{Fermi}-LAT data.
Src-N is significant above 5 GeV and spatially consistent with J1852 obtained in \cref{sec:lhaaso}.
However, with the accumulation of events, a new point source 4FGL~J1851.8-0007c (outside the range of the TS maps and not displayed) was added to 4FGL-DR4 catalog \citep{{2023arXiv230712546B}} and subsequently modeled as a Gaussian model correlated with SNR Kes~78 \citep{2025ApJ...993..115S}.
This new source is partially included by Src-S in terms of space and takes over a part of photons from that, which unavoidably affects the morphology of Src-S and subsequently interferes Src-N.
Therefore, we combined both the previous studies and reprocessed \textit{Fermi}-LAT observation with data for more than 16 years.
The gamma-ray source is considered to be significantly extended if ${\rm TS_{ext}=2\ln}(\cal L_{\rm ext}/L_{\rm ps})\geq$16 \citep{2012ApJ...756....5L}.

Taking into account the extension of 4FGL~J1851.8-0007c, we carefully analyzed the morphology of 4FGL~J1852.4+0037e by comparing the likelihood of two scenarios: (1) Refitting the location and extension of 4FGL~J1852.4+0037e. (2) Replacing 4FGL~J1852.4+0037e with two extended sources (SrcA$\rm _G$ and SrcB$\rm _G$).
The results show that the two extended source model (with 6 extra degrees of freedom) performs better with $\Delta$AIC$=-67.6$ and SrcA$\rm _G$ (with significance of $\sim16\sigma$) is spatially correspondent with the LHAASO source J1852 (see \cref{fig:tsmap_GeV}(d)).
Both SrcA$\rm _G$ and SrcB$\rm _G$ exhibit significant spatial extension, with TS$_{\rm ext}$ values of 46.1 and 61.6, respectively.
Due to the tiny discrepancy of Gaussian and Disk model for SrcA$\rm _G$ ($\Delta$TS$\le4$) we applied Gaussian model for SrcA$\rm _G$ similar to the morphology of the LHAASO source J1852.
The corresponding extension and position for the two GeV sources were shown in \cref{fig:tsmap_GeV}(d) and \cref{tab: spa_para_fermi}.
The LogP spectrums were adopted for these two sources with $E_0$ fixed to be 2262.93~MeV.
In addition, given the point-like gamma-ray emission overlapping with SNR Kes~79 (see \cref{fig:tsmap_GeV}(b)), we check another scenario building on the above by adding a point source at the position of SNR Kes~79 and then refitting the shape of SrcB$\rm _G$.
The resulted TS value $<25$ for this point source cannot claim a significant detection.
Thus, the possible contribution from SNR Kes~79 cannot be distinguished from SrcB$\rm _G$.

\begin{table*}
\caption{Results of the Spatial Analysis for \textit{Fermi}-LAT.}
\centering
\label{tab: spa_para_fermi}
\begin{tabular}{lccccc}
\hline\hline
\textbf{Model}                 & \textbf{R.A., decl. ($\degr$)} & \textbf{Extension ($\degr$)} & \textbf{Spectral parameters}  & \textbf{TS} \\
\hline
Gaussian (SrcA$\rm _G$)              & $283.40\pm 0.03\degr$, $1.10\pm 0.03 \degr$  &$0.19\pm 0.03\degr$ & $N_0=11.62\pm1.01$, $\varGamma = 2.70\pm0.14$,   & 313\\
                      &                                              &                              & $\beta=0.17\pm0.85$  & \\
Disk (SrcB$\rm _G$)              & $283.05\pm 0.04\degr$, $0.67\pm 0.03 \degr$  &$0.24\pm 0.02\degr$ & $N_0=8.28\pm0.89$, $\varGamma = 2.62\pm0.16$,  & 189  \\
                      &                                              &                              & $\beta=-0.03\pm0.08$ & \\
\hline
\end{tabular}
\begin{tablenotes}
\item Note: the values of $N_0$ are given in units of $10^{-13}$ MeV$^{-1}$cm$^{-2}$s$^{-1}$. The extension for Gaussian and Disk models refers to their respective 39\%-containment radii. The $1\sigma$ statistical uncertainties were given for morphological and spectral parameters. All parameters in this table were determined in the analysis of events with energies above $\sim$1 GeV.
\end{tablenotes}
\end{table*}

We also tested the other three spectral types PL, ECPL, and BPL for SrcA$\rm _G$ in separate fits.
The performance of these spectral types were shown in \cref{tab:spec}.
For SrcA$\rm _G$, the spectral type of LogP is the best and used in following analysis.

Based on the maximum likelihood analysis, the SED of source SrcA$\rm _G$ was produced by the \textit{sed} method in Fermipy in seven logarithmically spaced energy bins, as shown in \cref{fig:sed}. 
In this step, all event types (including PSF0, 1, 2, 3) were used to improve significance.
Same as in \cref{sec:lhaaso}, when fitting the flux points we only free the normalization of the sources with significance $\geq5\sigma$ and within 3$^{\circ}$ from the ROI centers, as well as the Galactic and isotropic diffuse background components.
For the energy bins with TS~$\le4$, the 95\% upper confidence limits were calculated.

In addition to statistic errors, we have taken into account the systematic uncertainty of spectrum. The systematic uncertainty on the Effective Area is about 5\% for analysis with energy dispersion correction disabled. The systematic uncertainty of spectrum from the choice of the Galactic interstellar emission model was simply estimated by artificially varying the normalization of the Galactic diffusion model by $\pm 6\%$ from the best-fit values in the entire energy band as well as in the individual energy bins of SED \citep{2009ApJ...706L...1A}. 
These two kinds of systematic errors and statistic errors were combined in quadrature.


\subsection{Molecular environment} \label{sec:mc}

\begin{figure*}
  \centering
	\includegraphics[width=7cm]{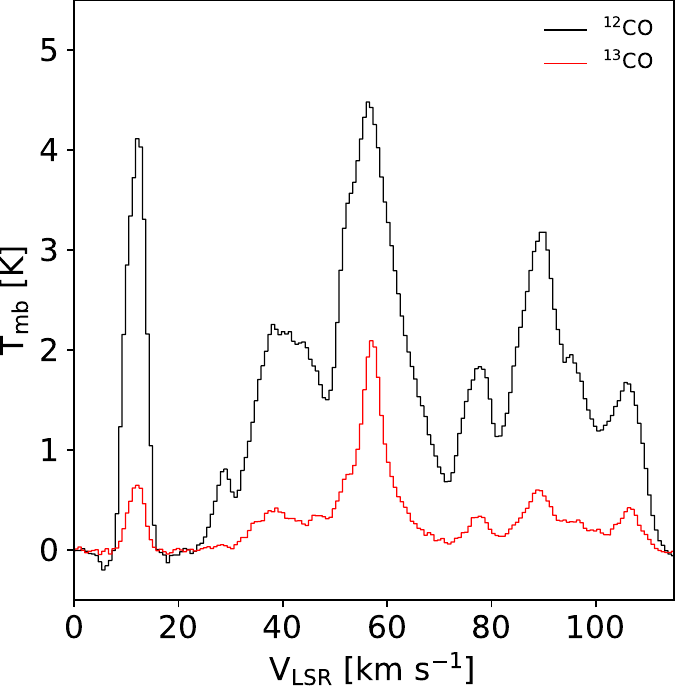}
  \caption{Spectra of $^{12}$CO J=1–0 and $^{13}$CO J=1–0 line emission in the SrcA${\rm_G}$ region.
  }
  \label{fig:COspec}
\end{figure*}

\begin{figure*}
  \centering
	\includegraphics[width=7cm]{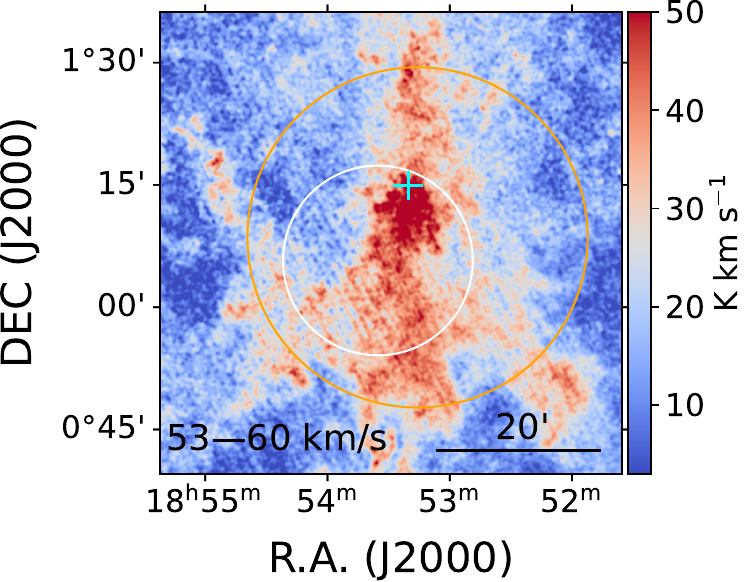}
	\includegraphics[width=7cm]{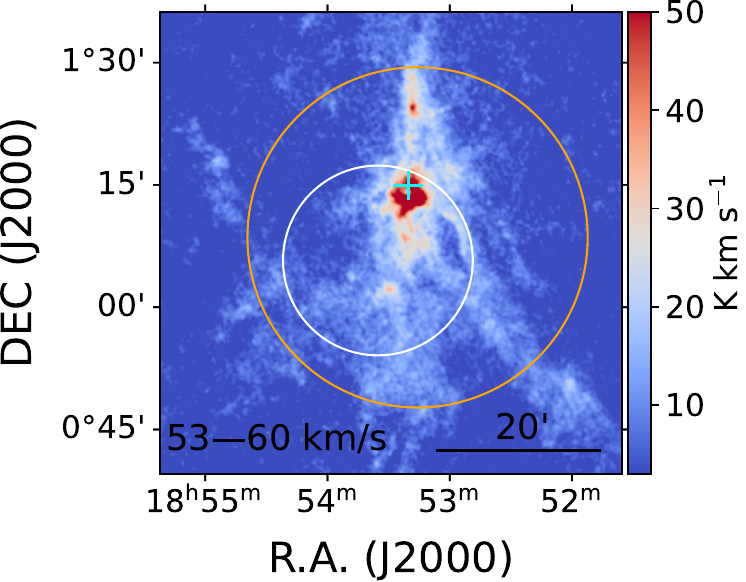}
  \caption{Intensity maps of \twCO\ \,(left) line and \thCO\ \,(right) line in the velocity range +53 to +60 km s$^{-1}$.
  The cyan plus represents the HII region G34.26+0.15.
  The white circles represent 39\% extension of the best-fitted spatial models for SrcA${\rm_G}$.
  The orange circle represents 1$\sigma$ Gaussian radius of J1852.
  }
  \label{fig:CO}
\end{figure*}

For a possibility of hadronic interaction, we studied the properties of the MC(s) coincident with SrcA${\rm_T}$ and SrcA${\rm_G}$.
The average main beam brightness temperature of the CO line emission (see Figure \ref{fig:COspec}) towards SrcA${\rm_G}$ within $\sim$11.4$^\prime$ from the center (see the white circle in \cref{fig:CO}) shows multiple peaks.
The MC in the local-standard-of-rest (LSR) velocity range of +53 -- +60 $\km\ps$ is the brightest component along the LOS and seems consistent with presence of the ultra-compact H\textsc{ii} region G34.26+0.15. 
\cref{fig:CO} shows the molecular emission integrated over the velocity range of +53 -- +60 $\km\ps$ around the detected gamma-ray sources, i.e.,
the LHAASO TeV source SrcA${\rm_T}$ (orange) and the GeV source SrcA${\rm_G}$ (white).
Within the spatial extent of the high-energy sources, there is a MC core, with the surrounding molecular gas extending predominantly along the north-south orientation.
The molecular core shown in the intensity maps of \thCO\ corresponds to the well-studied star-forming region G34.26+0.15 \citep[e.g.,][]{shock_evidence,NH3_maser}, which is marked with a cyan cross.

In this study, we use the FUGIN data to derive the physical parameters of the molecular gas in the region of high-energy sources.
Here, we calculate the column densities of molecular gas under these assumptions following \citet[][where detailed definitions of the symbols in the equations can be found]{Calculate_column_density} on the assumption that the \twCO\ emission is optically thick, while the \thCO\ emission is optically thin.
The column density of \thCO\ is given by
\begin{equation}
\begin{aligned}
    N_{^{13}\rm{CO}} & = \frac{3h}{8 \pi^3 \mu^2 J_\mu R_{\rm i}}\left(\frac{k T_{\rm ex}}{hB_{\rm 0}}+\frac{1}{3}\right) \exp{\frac{E_{\rm u}}{kT_{\rm ex}}} \\ & \times\left(\exp{\frac{h\nu}{kT_{\rm ex}}}-1\right)^{-1}\int\tau_{\rm \nu}dv.
	\label{equ:HCO+}
\end{aligned}
\end{equation}

On the basis of abundance ratio of $N({\rm H_2})/N({\rm ^{13}CO}) \approx 7 \times 10^5$ \citep{13CO-to-H2=7e5}, the resulting column densities $N({\rm H}_2)$ towards the TeV source J1852 and GeV source SrcA$_{\rm G}$ are 8.0$\times 10^{21}$~cm$^{-2}$ and 9.4$\times 10^{21}$~cm$^{-2}$, respectively, for molecular line profiles in the velocity range of +53 -- +60 $\km\ps$.
Assuming the LOS depth are comparable to the width of gamma-ray source region, we estimated a mean number density of H$_2$ molecules $\bar{n}({\rm H_2}) \sim$65 and 180 cm$^{-3}$ for J1852 and SrcA$_{\rm G}$, respectively.

\section{Discussion} \label{sec:dis}

The Fourth HAWC Catalog \citep{2026arXiv260200263A} reports a new TeV source, 4HWC J1854+0120, with a large extension of $\sim0.88\degr$. Its spatial extent encompasses the whole J1852 and partially overlaps with LHAASO J1857+0174.
However, given HAWC's angular resolution, the emission associated with J1852 cannot be reliably disentangled from that of the surrounding sources. Moreover, the absence of an H.E.S.S. Galactic plane survey \citep{2018A&A...612A...1H} counterpart is possibly due to the limited exposure time in this region and the complexity of source and diffuse-emission modeling in this crowded region.

With the SNR W44 and Kes~79 in the vicinity, as well as pulsar PSR~J1853+0056, dozens of H\textsc{ii} regions \citep{2014ApJS..212....1A}, and massive star-forming regions lying within the 39\%-containment radius $R_{39}$ (detailed in Appendix~\ref{App:a}), the TeV source J1852 is located at a complex projected region, which makes the exploration of the gamma-ray origin complicated.

Firstly, the physical relation of the gamma-ray sources J1852 and SrcA$_{\rm G}$ with the two SNRs can be excluded.
The distance to SNR W44 was estimated to be $\sim$2.5~kpc using the 1667~MHz OH absorption components at +42 $\km\ps$ \citep{1999ApJ...524..179C}, while the MC that may be associated with the ultra-compact HII region G34.26+0.15 and coincident with J1852 and SrcA$_{\rm G}$ (\cref{sec:mc}) is around 3.3--3.7~kpc (as mentioned in \cref{sec:intro}).
For SNR Kes~79, the escaped CRs is insufficient to afford the observed GeV gamma-ray flux \cite[the Src-N in ][]{2022ApJ...928...89H} under the canonical scenario \citep{2022ApJ...928...89H}.
Kes~79 was found to be interacting with the MC at the LSR velocity $\sim+80$ km s$^{-1}$ \citep{2018ApJ...864..161K};
at this velocity, we found no significant MC using the FUGIN observation within the $R_{39}$ of J1852 as a hadronic emitter.
Therefore, we proceeded to analyze the capability of SFRs and pulsar powering the observed GeV and TeV gamma-ray emission.

\cref{fig:sed} shows a smooth connection between the flux points of SrcA$\rm _G$ from \textit{Fermi}-LAT and J1852 from LHAASO.
However, the extension of J1852 is significantly larger than SrcA$\rm _G$, which may be indicative of different physical origins of the gamma-ray emissions around GeV and above TeV, or caused by energy-dependent diffusion.
We note that the first flux point of LHAASO data in \cref{fig:sed} yields only an upper limit, likely owing to the combined effects of source confusion in this very crowded region and the reduced sensitivity and broader point-spread function of WCDA at energies of a few TeV.

We suggest that SrcA$\rm _G$ is dominated by the jets of protostars in H\textsc{ii} regions but is not necessarily the GeV counterpart of J1852, with two cases discussed below. (1) The TeV emission (J1852) is dominated by the halo produced by pulsar PSR~J1853+0056; (2) The TeV emission (J1852) is dominated by the potential massive stars embedded in SFRs.

\subsection{Gamma-ray Contribution from Pulsar Halo}\label{subsec:halo_TeV}
As PWNe evolve, the energetic positrons and electrons accelerated internally can diffuse into the ambient ISM.
During the diffusion process, electrons are capable of radiating gamma-ray photons by inverse Compton (IC) scatterings and form extended gamma-ray source, called pulsar halo, in TeV sky.
The most prominent pulsar halos are Geminga and Monogem reported by High Altitude Water Cherenkov Observatory \citep{2017Sci...358..911A}.
In this scenario, the pulsar PSR~J1853+0056 with spin-down luminosity of $4.0\times10^{34}$ erg s$^{-1}$, characteristic age of $\tau_{\rm c}\sim2\times10^5$~yr, and DM-based distance of 3.8~kpc \citep{2002MNRAS.335..275M} could be expected to contribute to the TeV emission of J1852.

\begin{figure*}
  \centering
	\includegraphics[width=8cm]{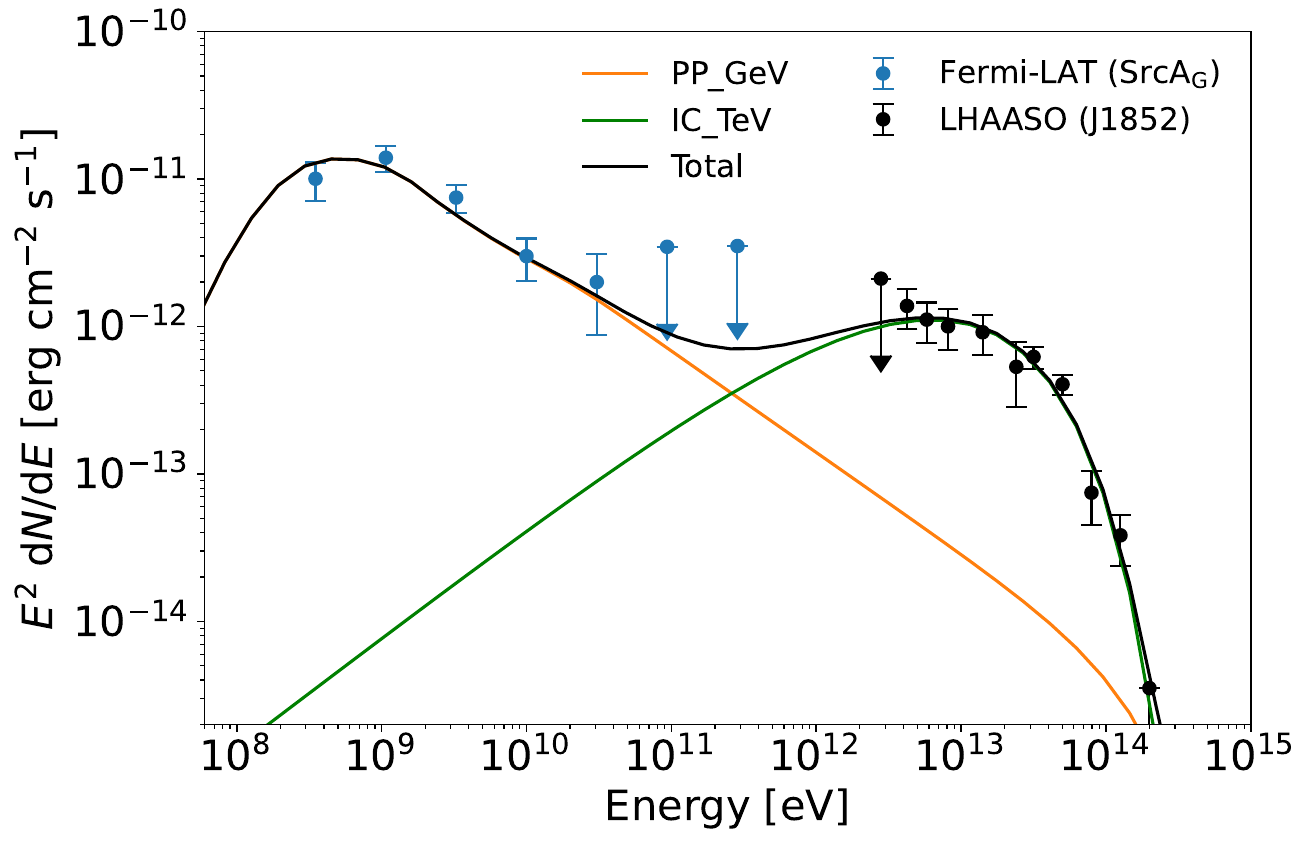}
	\includegraphics[width=8cm]{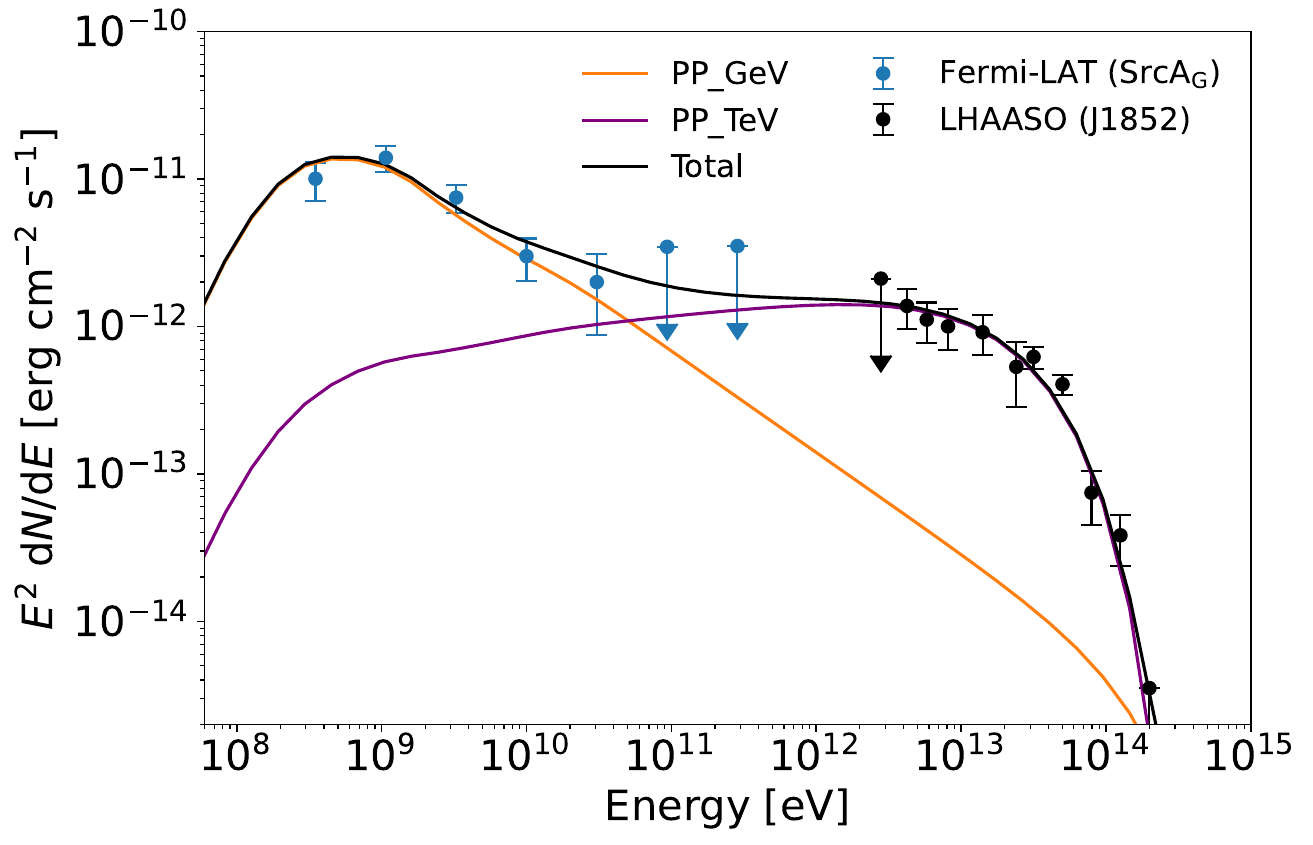}
  \caption{SED fitting of GeV-100TeV gamma-rays of SrcA$_{\rm G}$ and J1852 in 1LHAASO J1852+0050u region. The orange lines represent the hadronic contribution of the protons accelerated by protostar activities. The green line in the left panel represents the IC emission from the electrons of pulsar(s). The purple line in the right panel represents the hadronic contribution from the massive stars in SFR.
  }
  \label{fig:sed_fit1}
\end{figure*}

We first used the python package Naima\footnote{\url{https://naima.readthedocs.io/en/latest/mcmc.html}} (version 0.10.0) to fit the SED in TeV band.
We simply assumed an ECPL distribution for whatever positrons or electrons accelerated by PSR J1853+0056 or hidden pulsar(s) to fit the flux points of J1852 (shown with green curve in the left panel of \cref{fig:sed_fit1}).
Applying the analytical approximations developed by \citep{2014ApJ...783..100K}, the resulted power-law index, cutoff, and total particles energy above 1 GeV are $\varGamma=1.5$, $E_{\rm cutoff}=35$ TeV, and $W_e=2.2\times10^{46}$~erg, respectively.
The seed photon field for the relativistic electrons to scatter includes cosmic microwave background, far-infrared dust emission, and near-infrared stellar emission with $[T_{\rm CMB},\ T_{\rm FIR},\ T_{\rm NIR}]=[2.72,\ 30,\ 3000]\ {\rm K}$ and $[u_{\rm CMB},\ u_{\rm FIR},\ u_{\rm NIR}]=[0.26,\ 0.5,\ 1]\ {\rm eV\ cm^{-3}}$.

In this case, we fit the SED of SrcA$\rm _G$ with a hadronic model by assuming a PL distribution for protons.
The resulted proton index is $\varGamma\sim2.75$ and the required total energy of the protons above 1 GeV for SrcA$\rm _G$ is $W_p\approx2.1\times10^{48}(d/\rm 3.3\ kpc)^2(n_{\rm t} /360\ {\rm cm^{-3}})^{-1}$~erg, where the $d$ and $n_{\rm t}$ is the distance to the GeV gamma-ray source and the number density of atomic hydrogen.

For analyzing the ability of accelerating particles, then we used an one-zone and time-dependent diffusion model (detailed in Appendix~\ref{App:b}) to fit the TeV emission.
As seen in \cref{fig:sed_halo}, by varying the adjustable parameters including index, break energy of the electrons distribution, and the age of PSR~J1853+0056, we found that the flux of halo model for any parameters cannot reach the observed flux of TeV emission even when a relatively low magnetic field (3$\mu$G) was adopted.
Due to the low spin-down power and the cooling from synchrotron and IC, the fitted results exhibit a small flux in TeV band.
Thus, PSR~J1853+0056 seems not the dominant origin of the TeV source J1852 and there should be other objects to accelerate high-energy particles.
However, we cannot rule out the contribution from hidden pulsars
in this region with pulse beam away from the Earth.

\subsection{Gamma-ray Contribution from H\textsc{ii} Regions}
\subsubsection{Protostars dominate the GeV emission} \label{subsec:Proto_GeV}

 
Our CO observation analysis shows that the MC that is correlated with the ultra-compact H\textsc{ii} region G34.26+0.15 is well spatially coincident with the gamma-ray source SrcA$\rm _G$ and J1852.
In addition to the massive stars which ionize the H\textsc{ii} region, there may be other embedded young stars.
Moreover, the explosive dispersal outflow, which differs from the protostellar jets driven by individual young stars, has been identified in this H\textsc{ii} region \citep{2025AJ....169..324I}.
Such an explosive outflow is possibly triggered by the young, massive, nonhierarchical stellar system, potentially triggered by the merger of massive protostars \citep{2009ApJ...704L..45Z,2017ApJ...836..133Z}.
The GeV gamma-ray emission potentially associated with DR21, G34.26+0.15, and G5.89$-$0.39 has recently been reported and connected with explosive outflow \citep{2026ApJ..1000..102P}.
Moreover, \cite{2026NatAs.tmp..133M} report a statistically significant detection of gamma-rays from a population of young stellar objects.

Therefore, we suggest that the GeV emission (SrcA$\rm _G$) observed by \textit{Fermi}-LAT arises from the MC illuminated by the CRs accelerated by protostars's activities, such as outflows or jets.
Assuming a PL distribution for the accelerated protons, the resulting hadronic spectrum for SrcA$\rm _G$ is shown in \cref{fig:sed_fit1} (orange curve in both panel).
By adopting the distance to G34.26+0.15 as 3.3 kpc obtained from H\textsc{i} absorption studies \citep{1994ApJ...436..117K}
and the atomic number density $\bar{n}({\rm H}) \sim$360 cm$^{-3}$ of the MC in the region of SrcA$\rm _G$ region (obtained in \cref{sec:mc}), the fitted proton index is $\varGamma\sim2.75$ and the required total energy of the protons for SrcA$\rm _G$ is $W_p\approx2.1\times10^{48}$~erg.
It is noteworthy that in this case we used the same population of protons as in \cref{subsec:halo_TeV}.
The fitted proton index $\varGamma\sim2.75$ is similar to that obtained in HH 80-81,  the most powerful protostellar jets in the Galaxy \citep{2022RAA....22b5016Y,2025A&A...695A..11M}.





\subsubsection{Massive stars dominate the TeV emission} \label{subsec:HII_TeV}

Similar to the case in G34.26+0.15C \citep{2025AJ....169..324I} (see \cref{sec:intro}), an explosive outflow was identified in SFR G5.89$-$0.39 from the molecular gas kinematics.
G5.89$-$0.39 overlaps with TeV source HESS~J1800-240B projectively and is believed to account for a part of the GeV--TeV gamma-ray emission \citep{2016JHEAp..11....1H}.
Therefore, we expect that G34.26+0.15 could also emit TeV gamma-rays (J1852) in the light of the potential for massive stars in SFRs to operate as PeVatrons.

As shown by the purple curve in the right panel of \cref{fig:sed_fit1}, an ECPL distribution of protons can fit the TeV flux points of J1852 well.
The index, cutoff, and total energy of protons are $\varGamma\sim2.0$, $E_{\rm cutoff}=250$ TeV, and $W_p=4.3\times10^{47}$~erg, respectively.
This proton energy $W_p$ for J1852 is almost one order of magnitude lower than that for SrcA$\rm _G$ (see \cref{subsec:Proto_GeV}). 
The index $\varGamma\sim2.0$ is consistent with the prediction of index $\le2.3$ for the particle acceleration in stellar clusters \citep{2019NatAs...3..561A}.

We searched for the OB stars in the catalog of LAMOST DR11 \citep{2012RAA....12..723Z}, VizieR \citep{2014yCat....1.2023S}, and \textit{Gaia} DR3 database \citep{2023A&A...674A...1G,2023A&A...674A..39G}.
The results show that most of the OB stars in the 39\% containment radius $R_{39}$ of J1852 are within 3~kpc of the Earth and no one overlaps with G34.26+0.15.
In fact, considering the extinction by dense gas in these ultra-compact H\textsc{ii} regions, we are not surprised by the non-detection of OB stars and cannot rule out that they are embedded deep within the MC.
The number of the ultra-compact H\textsc{ii} regions in the Galaxy requires multiple embedded massive stars to sustain by stellar wind or radiation pressure \citep{1989ApJS...69..831W}.
In addition, we also note that two stellar clusters, [BDS2003] 128 and [BDS2003] 127, in the catalog of \cite{2013A&A...560A..76M} are correlated with G34.26+0.15.

Thus, we postulate that the energetic protons in TeV band are accelerated by the stellar winds of massive stars and discuss this possibility from the energy budget side.
Exploiting the method in \cite{2025ApJ...979...70C}, the proton luminosity $L_{\rm p}(>1\ {\rm GeV})\sim0.2$--$2.6\times 10^{36}$~erg~s$^{-1}$ for $\delta =0.3-0.6$ was roughly derived from $\int_{\rm 1 GeV}^{+\infty} E_{\rm p} {\rm d}N_{\rm p}/{\rm d}E_{\rm p}/{\rm min}(\tau_{\rm pp},t_{\rm c}) {\rm d}E_{\rm p}$, where the $\tau_{\rm pp}$ and $t_{\rm c}$ are proton's lifetime and confinement time, respectively. 
The lifetime was estimated using $\tau_{\rm pp}=10^{5} (n_{\rm t} /600\ {\rm cm^{-3}})^{-1}$yr with the atomic hydrogen number density $n_{\rm t}=130$ cm$^{-3}$ in the region of J1852 obtained in \cref{sec:mc}.
The confinement time is dominated by diffusion and given by $t_{\rm c}=l^2/6D$, where $l$ and $D$ denote the physical size of gamma-ray source and diffusion coefficient, respectively.
At the distance of $\sim3.3$~kpc, the physical size of J1852 was estimated to be $\sim30$~pc corresponding to the 68\%-containment (that is, 1.51$R_{39}$ for the two-dimensional Gaussian).
We assumed the energy-dependent diffusion coefficient following $D=\chi 10^{28}(E_{\rm p}/10\ {\rm GeV})^{\delta}\ {\rm cm^2\ s^{-1}}$.
By adopting the same diffusion factor $\chi=0.01$ and mechanical power in wind of a typical OB-type star $\sim 10^{35}\ {\rm erg\ s^{-1}}$ as in \cite{2025ApJ...979...70C}, the number of massive stars $N\sim 20$--$260$ for $\delta =0.3-0.6$ is needed.
Here, about 10\% of the total energy in wind were reasonably assumed to accelerate protons.
The diffusion index $\delta$ was varied in terms of different diffusive processes.
This number may be further reduced due to the magnetic turbulence amplified by multiple H\textsc{ii} regions, accounting for a significant suppression to diffusive process.

Considering the perfect coincidence between the MC (corresponding to the H\textsc{ii} region G34.26+0.15) and the gamma-ray emission and lack of a pulsar powering the TeV emission, we suggest that G34.26+0.15 with evident star-forming activities is the most possible accelerator responsible for the observed gamma-ray emission.

\subsubsection{The difference in the extension of SrcA$\rm _G$ and J1852}

Considering that both the protostars and massive stars are located in SFR G34.26+0.15, and that the size of SFR G34.26+0.15 is smaller than gamma-ray sources, the CR acceleration region responsible for the GeV and TeV emission should be the same and can be treated as a point source.
In this case, the difference in the extensions of SrcA$\rm _G$ and J1852 may be attributed to the energy-dependent diffusion.
As in \cref{fig:CO}, the extension of MC is comparable to that of J1852 but larger than that of SrcA$\rm _G$.
This suggests that the low-energy CRs cannot escape from the MC during their lifetimes for such a continuous injection scenario.
The \textit{Fermi}-LAT SED shows no valid flux point above $\sim$30 GeV.
By assuming that the lifetime is comparable to the confinement time for CRs with energy below 30 GeV, the diffusion factor is constrained to be $\sim$0.004 (for a diffusion index $\delta=0.5$).
This value is consistent with the diffusion factor assumed in \cref{subsec:HII_TeV}.

\subsubsection{The possibility that massive stars dominate the GeV and TeV gamma-ray emission}

We also examined the possibility that the CRs responsible for the GeV and TeV gamma-ray emission belong to the same population accelerated by massive stars.
Since the low-energy CRs are expected to remain confined within the MC, the proton index $\varGamma\sim2.75$ derived from phenomenological fit to the GeV emission (see \cref{subsec:Proto_GeV}) should approximately reflect the intrinsic index of accelerated particles.
However, the broadband gamma-ray emission from GeV to TeV band cannot be reproduced by such a proton distribution.
The Cygnus OB2 exhibits a similar gamma-ray SED \citep{2024SciBu..69..449L}.
The study of Cygnus OB2 places the molecular cloud at distance of 100 pc from the accelerator in order to reproduce the GeV-TeV emission simultaneously using a single population of protons. However, the $^{13}$CO line in our study peaks at the position of G34.26+0.15 (see \cref{fig:CO}), indicating that the molecular cloud should be coincident with this star-forming region in space. Therefore, the diffusion-driven spectrum invoked for Cygnus OB2 is not expected here. In this case, the steep GeV emission and flat-to-steep TeV emission cannot be explained simultaneously.

\section{Conclusion}  \label{sec:concl}

With the accumulation of LHAASO data, we reanalyzed the extended TeV source 1LHAASO~J1852+0050u (with the refitted source LHAASO~J1852+0050u called J1852 in this study).
Using $\sim$4 years of WCDA and KM2A data, the J1852, with a significance of $\sim13\sigma$ above 1 TeV, was found in the southwest of the SNR W44 and overlapped with dozens of H\textsc{ii} regions and the PSR~J1853+0056.

To explore the origin of J1852, we combined previous studies and reprocessed $\sim16$ years of \textit{Fermi}-LAT.
We derived an extended GeV source (SrcA$\rm _G$), which was denoted to coincide with the LHAASO source of J1852, with a significance of $\sim16\sigma$ in 1--500 GeV. However, the difference in extension between these two gamma-ray sources may hint different specific physical origins.

The $^{12}$CO and $^{13}$CO data from Nobeyam show a molecular clump in the velocity of +53 -- +60 km s$^{-1}$ correlated with the H\textsc{ii} regions G34.26+0.15.
Incorporating the LHAASO and \textit{Fermi}-LAT data, this MC appears coincident with the TeV and GeV gamma-ray emission spatially.

Thus, we fitted the SED of J1852 and SrcA$\rm _G$ simultaneously with two scenarios: (1) The TeV emission (J1852) is dominated by the halo produced by pulsar PSR~J1853+0056; (2) The TeV emission (J1852) is dominated by the potential massive stars.
In the first scenario, we used an one-zone and time-dependent diffusion model to simulate the gamma-ray emission of pulsar halo from the PSR~J1853+0056.
For the parameter spaces as reasonable as possible, the PSR~J1853+0056 seems not enough to power such a TeV halo which is responsible for the observed flux of TeV emission.
For the second scenario, considering the soft index of SrcA$\rm _G$ in the \textit{Fermi}-LAT data and the existence of abundant star-forming regions, we attributed this GeV emission to the possible protostars and discussed the massive stars embedded in H\textsc{ii} regions dominating the TeV emission.
Using an ECPL distribution of protons to fit the TeV flux points, the resulted proton index of $\varGamma\sim2.0$ is consistent with other gamma-ray sources associated stellar clusters.
Given the required proton energy $W_p\approx4.3\times10^{47}(n_{\rm t}/130{\rm\,cm}^{-3})^{-1}$~erg, the number of massive stars $N \sim 20$--$260$ for $\delta =0.3-0.6$ are needed to sustain the observed TeV emission.
Thus, we suggest that the GeV and TeV emissions arise from the MCs illuminated by the CRs from the jets of protostars and massive stars in H\textsc{ii} regions, respectively.
%

\section{Acknowledgements}
This publication makes use of data from FUGIN, FOREST Unbiased Galactic plane Imaging survey with the Nobeyama 45-m telescope, a legacy project in the Nobeyama 45-m radio telescope. This work has made use of data from the European Space Agency (ESA) mission Gaia (\url{https://www. cosmos.esa.int/gaia}), processed by the Gaia Data Processing and Analysis Consortium (DPAC, \url{https://www. cosmos.esa.int/web/gaia/dpac/consortium}). Funding for the DPAC has been provided by national institutions, in particular the institutions participating in the Gaia Multilateral Agreement. We would like to thank the staff members of National Facility for Large High Altitude Air Shower Observatory (LHAASO, \url{https://cstr.cn/31117.02.LHAASO}) who work at the LHAASO site above 4400 meter above the sea level year round to maintain the detector and keep the water recycling system, electricity power supply and other components of the experiment operating smoothly. We are grateful to Chengdu Management Committee of Tianfu New Area for the constant financial support for research with LHAASO data. We appreciate the computing and data service support provided by the National High Energy Physics Data Center for the data analysis in this paper. 
This work is partially supported by National Natural Science
Foundation of China (NSFC) under grants 12121003, 12393852, and 12573047 and the Basic Research Program of Jiangsu NO.BK20252108. 
It is also supported by the NSFC under grants 
12393851, 12393852, 12393853, 12393854, 12205314, 12105301, 12305120, 12261160362, 12105294, U1931201, 12375107, and 12173039,
the Department of Science and Technology of Sichuan Province, China No.24NSFSC2319, Project for Young Scientists in Basic Research of Chinese Academy of Sciences No.YSBR-061, and in Thailand by the National Science and Technology Development Agency (NSTDA) and the National Research Council of Thailand (NRCT) under the High-Potential Research Team Grant Program (N42A650868).

\section{Author contributions}
Y.\ Chen organised this study; C.\ Huang analysed the data of the LHAASO WCDA and KM2A, \textit{Fermi}-LAT, CO-line, and radio observations; Y.Z.\ Shen investigated the optical survey data, X.\ Zhang, B.\ Liu, and Y.\ Chen led the physical interpretation; C.\ Huang and J.X.\ Sun performed the model calculation of pulsar halo; C. Huang, X.\ Zhang, B.\ Liu, Y.\ Chen, and Y.Z.\ Shen prepared the manuscript. S.C.\ Hu and M.\ Zha provided the pipeline for the joint analysis. Y.Z.\ Shen provided the crosscheck. All other authors participated in data analysis, including detector calibration, data processing, event reconstruction, data quality check, and various simulations, and provided comments on the manuscript.

\section*{Data Availability}
The \emph{Fermi}-LAT data underlying this work are publicly available, and can be downloaded from \url{https://fermi.gsfc.nasa.gov/ssc/data/access/lat/}.
The FUGIN data are publicly available in \url{https://jvo.nao.ac.jp/portal/nobeyama/fugin.do}.
The SMGPS data are publicly available, and provided by \url{https://archive-gw-1.kat.ac.za/public/repository/10.48479/3wfd-e270/index.html}.
The LHAASO data underlying this article will be shared on reasonable request to the corresponding author.

\bibliographystyle{mnras}
\bibliography{reference}

\par
\addvspace{\baselineskip}
{\small\itshape\noindent
$^{1}$ State Key Laboratory of Particle Astrophysics \& Experimental Physics Division \& Computing Center, Institute of High Energy Physics, Chinese Academy of Sciences, 100049 Beijing, China\\
$^{2}$ University of Chinese Academy of Sciences, 100049 Beijing, China\\
$^{3}$ TIANFU Cosmic Ray Research Center, 610000 Chengdu, Sichuan,  China\\
$^{4}$ University of Science and Technology of China, 230026 Hefei, Anhui, China\\
$^{5}$ Yerevan State University, 1 Alek Manukyan Street, Yerevan 0025, Armeni a\\
$^{6}$ Max-Planck-Institut for Nuclear Physics, P.O. Box 103980, 69029  Heidelberg, Germany\\
$^{7}$ Tsung-Dao Lee Institute \& School of Physics and Astronomy, Shanghai Jiao Tong University, 200240 Shanghai, China\\
$^{8}$ Center for Astrophysics, Guangzhou University, 510006 Guangzhou, Guangdong, China\\
$^{9}$ APC, Universit\'e Paris Cit\'e, CNRS/IN2P3, CEA/IRFU, Observatoire de Paris, 119 75205 Paris, France\\
$^{10}$ Institute for Nuclear Research of Russian Academy of Sciences, 117312 Moscow, Russia\\
$^{11}$ School of Physical Science and Technology \&  School of Information Science and Technology, Southwest Jiaotong University, 610031 Chengdu, Sichuan, China\\
$^{12}$ Department of Physics, The Chinese University of Hong Kong, Shatin, New Territories, Hong Kong, China\\
$^{13}$ State Key Laboratory of Particle Detection and Electronics, China\\
$^{14}$ Key Laboratory of Dark Matter and Space Astronomy \& Key Laboratory of Radio Astronomy, Purple Mountain Observatory, Chinese Academy of Sciences, 210023 Nanjing, Jiangsu, China\\
$^{15}$ Hebei Normal University, 050024 Shijiazhuang, Hebei, China\\
$^{16}$ Shanghai Astronomical Observatory, Chinese Academy of Sciences, 200030 Shanghai, China\\
$^{17}$ School of Physics and Astronomy, Yunnan University, 650091 Kunming, Yunnan, China\\
$^{18}$ Key Laboratory of Cosmic Rays (Tibet University), Ministry of Education, 850000 Lhasa, Tibet, China\\
$^{19}$ School of Astronomy and Space Science, Nanjing University, 210023 Nanjing, Jiangsu, China\\
$^{20}$ Key Laboratory of Radio Astronomy and Technology, National Astronomical Observatories, Chinese Academy of Sciences, 100101 Beijing, China\\
$^{21}$ School of Physics and Astronomy (Zhuhai) \& School of Physics (Guangzhou) \& Sino-French Institute of Nuclear Engineering and Technology (Zhuhai), Sun Yat-sen University, 519000 Zhuhai \& 510275 Guangzhou, Guangdong, China\\
$^{22}$ School of Physics and Electronic Science, Guizhou Normal University, 550025 Guiyang, China\\
$^{23}$ Research Center for Astronomical Computing, Zhejiang Laboratory, 311121 Hangzhou, Zhejiang, China\\
$^{24}$ Institute of Frontier and Interdisciplinary Science, Shandong University, 266237 Qingdao, Shandong, China\\
$^{25}$ Department of Engineering Physics \& Department of Physics \& Department of Astronomy, Tsinghua University, 100084 Beijing, China\\
$^{26}$ Yunnan Observatories, Chinese Academy of Sciences, 650216 Kunming, Yunnan, China\\
$^{27}$ China Center of Advanced Science and Technology, Beijing 100190, China\\
$^{28}$ College of Physics, Sichuan University, 610065 Chengdu, Sichuan, China\\
$^{29}$ Center for Relativistic Astrophysics and High Energy Physics, School of Physics and Materials Science \& Institute of Space Science and Technology, Nanchang University, 330031 Nanchang, Jiangxi, China\\
$^{30}$ School of Physics \& Kavli Institute for Astronomy and Astrophysics, Peking University, 100871 Beijing, China\\
$^{31}$ Guangxi Key Laboratory for Relativistic Astrophysics, School of Physical Science and Technology, Guangxi University, 530004 Nanning, Guangxi, China\\
$^{32}$ Department of Physics, Faculty of Science, Mahidol University, Bangkok 10400, Thailand\\
$^{33}$ School of Physics and Technology, Nanjing Normal University, 210023 Nanjing, Jiangsu, China\\
$^{34}$ Moscow Institute of Physics and Technology, 141700 Moscow, Russia\\
$^{35}$ National Space Science Center, Chinese Academy of Sciences, 100190 Beijing, China\\
$^{36}$ School of Physics, Huazhong University of Science and Technology, Wuhan 430074, Hubei, China\\
}

\appendix 
\section{Supplementary Material} \label{App:a}

41 of H\textsc{ii} regions, two stellar clusters, and a high-energy pulsar within the 39\%-containment radius of J1852 were listed in \cref{tab:HII}.
Among these H\textsc{ii} regions, NRAO584 (G034.254+00.144, \citealt{2014ApJS..212....1A}) is very bright in infrared and radio bands and is composed of five H\textsc{ii} regions.
These five H\textsc{ii} regions, which contain G034.236+00.166, G034.246+00.102, G034.256+00.136, G034.256+00.154, and G034.295+00.089, were grouped according to the spatial intersection of the photo-dissociation region.
The combination of G034.256+00.136 and G034.256+00.154 is also called G34.26+0.15.

\begin{table*}
\centering
\caption{The high-energy objects within the $R_{39}$ of J1852}
\begin{tabular}{m{2cm}m{1cm}m{3cm}m{5.7cm}}
\hline\hline
 & \textbf{Type$^a$} & \textbf{Group} & \textbf{Name} \\\hline
\multirow{8}{*}{H\textsc{ii} regions} & K & & G033.882+00.057$^1$, G033.884+00.058$^1$, G033.914+00.107$^3$, G033.941$-$00.039$^4$, G033.987$-$00.012$^1$, G033.991$-$00.005$^1$, G034.026$-$00.058$^1$, G034.041+00.052$^1$, G034.089+00.438$^6$, G034.333+00.212$^1$, G034.404+00.227$^6$, G034.443+00.103$^8$ \\\cline{2-4}
 & Q & & G033.894$-$00.078$^2$, G034.023+00.233$^2$, G034.251$-$00.120$^2$, G034.259+00.015$^2$, G034.269$-$00.208$^2$, G034.298+00.180$^2$, G034.316+00.009$^2$, G034.428$-$00.073$^2$, G034.429$-$00.053$^2$, G034.443+00.056$^2$, G034.484+00.013$^2$ \\\cline{2-4}
 & C & & G034.095+00.017$^2$, G034.130$-$00.174$^2$, G034.174$-$00.086$^2$, G034.190$-$00.063$^2$, G034.469$-$00.020$^2$ \\\cline{2-4}
 &\multirow{5}{*}{G} & G034.047+00.141 & G034.046+00.140$^5$, G034.051+00.044$^2$\\\cline{3-4}
 &  & G034.105$-$00.046 & G034.102$-$00.027$^5$, G034.104$-$00.046$^2$ \\\cline{3-4}
 &  & G034.172+00.175 & G034.137+00.094$^2$, G034.158+00.147$^1$ \\\cline{3-4}
 &  & G034.172+00.175 & G034.201+00.105$^2$, G034.214+00.119$^2$ \\\cline{3-4}
 &  & G034.254+00.144 & G034.236+00.166$^2$, G034.246+00.102$^2$, G034.256+00.136$^7$, G034.256+00.154$^2$, G034.295+00.089$^2$ \\\hline
 \multirow{2}{*}{Stellar clusters} & & & [BDS2003] 128$^9$ \\
  & & & [BDS2003] 127$^9$ \\\hline
  Pulsar & & & PSR J1853+0056$^{10}$\\\hline
\end{tabular}
\begin{flushleft}
\footnotesize
\item $^a$ The label named K, G, C, and Q represent the known, group, candidate, and radio quiet sources identified in \cite{2014ApJS..212....1A}.
\item Reference: (1) \cite{2011ApJS..194...32A}; (2) \cite{2014ApJS..212....1A}; (3) \cite{2002ApJS..138...63A}; (4) \cite{1996ApJ...472..173L}; (5) \cite{2015ApJS..221...26A}; (6) \cite{2003ApJ...587..714W}; (7) \cite{2011ApJ...738...27B}; (8) \cite{2018ApJS..234...33A}; (9) \cite{2013A&A...560A..76M}; (10) \cite{2002MNRAS.335..275M}.
\label{tab:HII}
\end{flushleft}
\end{table*}

\section{Pulsar halo model} \label{App:b}

The propagation of electrons can be described by the diffusion equation \citep{1964ocr..book.....G}
\begin{equation}
\frac{\partial}{\partial t}f(\gamma, r, t) = \frac{D(\gamma)}{r^2}\frac{\partial}{\partial r}r^2\frac{\partial}{\partial r}f(\gamma, r, t) + \frac{\partial}{\partial \gamma}(P f) + Q(\gamma, t)\delta(r).
\label{eq:diff}
\end{equation}
The $f(\gamma, r, t)$ denotes the energy distribution function of electrons at time $t$ and distance $r$ from the source.
The $D(\gamma)$ is the energy-dependent diffusion coefficient.
The second and third terms in the right side represent the continuous change due to energy losses and injection.
Here, we assumed that the injection rate followed a broken power-law 
\begin{equation}
Q(\gamma, t)=Q_0(t)
\begin{cases}
 \left(\frac{\gamma}{\gamma_{\rm b}}\right)^{-\alpha_1},\gamma\le \gamma_{\rm b} \\ 
 \left(\frac{\gamma}{\gamma_{\rm b}}\right)^{-\alpha_2},\gamma\ge \gamma_{\rm b}
\end{cases},
\end{equation}
where the $\gamma_{\rm b}$ is the break energy.
The energy-dependent normalization $Q_0(t)$ is determined based on the injection luminosity of pulsar \citep[e.g.,][]{2012MNRAS.427..415M,2023A&A...673A.148H}
\begin{equation}
L(t)=\int_{0}^{\infty}\gamma mc^2Q(\gamma, t)d\gamma.
\end{equation}
The time-dependent evolution of injection luminosity is
\begin{equation}
L(t)=L_0\left(1+\frac{t}{\tau_0}\right)^{-\frac{n+1}{n-1}},
\end{equation}
where the $L_0$ and $\tau_0$ denote the initial luminosity and the initial spin-down time-scale of the pulsar, with $n$ the breaking index.
The $\tau_0$ is given by \citep{2006ARA&A..44...17G}
\begin{equation}
\tau_0=\frac{2\tau_{\rm c}}{n-1}-t_{\rm age},
\label{eq:tau0}
\end{equation}
where the $\tau_{\rm c}$ is the characteristic age of the pulsar obtained by observation.
For a given $t_{\rm age}$, the initial luminosity $L_0$ can be determined according to the luminosity $L(t_{\rm age})=4.0\times10^{34}$~erg~s$^{-1}$.

We obtained the solution of \cref{eq:diff} for the continued injection of electrons \citep{1995PhRvD..52.3265A,2019PhRvD.100l3015D}:
\begin{equation}
\begin{aligned}
&f(\gamma, r, t)\\
&=\int_{0}^tdt_0 \frac{Q(\gamma_t(\gamma,t_0,t),t_0)P(\gamma_t(\gamma,t_0,t))} {\pi^{3/2}P(\gamma)r_{\rm dif}^3}{\rm exp}\left(-\frac{r^2}{r_{\rm dif}^2}\right)\\
&=\int_{0}^tdt_0\frac{Q(\gamma_t(\gamma,t_0,t),t_0)}{\pi^{3/2}r_{\rm dif}^3} [1-p_2(t-t_0)\gamma]^{-2} {\rm exp}\left(-\frac{r^2}{r_{\rm dif}^2}\right),
\end{aligned}
\end{equation}
where the $\gamma_t(\gamma,t_0,t)=\gamma/(1-p_2(t-t_0)\gamma)$ was defined as the initial energy of electrons which were produced at time $t_0$ and cooled down to $\gamma$ after time $t-t_0$, with $p_2$ the energy-loss coefficient of synchrotron and IC.
The effective diffusion radius
\begin{equation}
r_{\rm dif}(\gamma, t_0, t)\simeq 2\sqrt{D(\gamma)(t-t_0)\frac{1-(1-\gamma/\gamma_{\rm cut})^{(1-\delta)}} {(1-\delta)\gamma/\gamma_{\rm cut}}},
\end{equation}
where $\gamma_{\rm cut}=[p_2(t-t_0)]^{-1}$, corresponding to the maximum radius which electrons with energy $\gamma$ can reach during the time $t-t_0$ after their injection from the source \citep{1995PhRvD..52.3265A}.
The $\delta$ is the diffusion index.
\cref{tab:halo} shows the parameters from observation and variables in the model.

\begin{table*}
\centering
\caption{Model parameters.}
\begin{tabular}{lllr}
\hline\hline
\textbf{Parameters} & \textbf{Description} & \textbf{Value} & Ref.\\
\hline
\textit{d} (kpc)        & distance to the pulsar    & 3.8 & (1)\\
$L$ (erg s$^{-1}$)      & pulsar spin-down power    & $4\times10^{34}$ & (1)\\
$\tau_{\rm c}$ (kyr)    & pulsar characteristic age & $2\times10^2$ & (1)\\
$P$ (ms)                & pulsar period             & 275 & (1)\\
$\dot P$ (s s$^{-1}$)   & pulsar period derivative  & 2.139$\times10^{-14}$ & (1) \\
$n$                     & pulsar braking index      & 3 & fixed\\
$\delta$                & diffusion index           & 0.33 & fixed\\
$D_{100}$ (cm$^2$ s$^{-1}$) & diffusion coefficient for particles in 100 TeV & $3\times10^{27}$& fixed\\
$B_{\rm halo}$ ($\mu$G) & field strength in the TeV halo & 3& fixed\\
$T_{\rm NIR}$ (K)       & NIR temperature           & 3000 & fixed\\
$u_{\rm NIR}$ (eV cm$ ^{-3}$) & NIR energy density  &  1  & fixed\\
$T_{\rm FIR}$ (K)       & FIR temperature           & 30  & fixed\\
$u_{\rm FIR}$ (eV cm$ ^{-3}$) & FIR energy density  & 0.5  & fixed\\
$T_{\rm CMB}$ (K)       & CMB temperature           & 2.72  & fixed\\
$u_{\rm CMB}$ (eV cm$ ^{-3}$) & CMB energy density  &  0.26  & fixed\\
$\alpha_1$              & index of electrons in BPL & 1.6 & fixed\\
$\alpha_2$              & index of electrons in BPL & [2.2, 2.7, 3.2] & fitted\\
$\gamma_{\rm b}$        & break energy              & [$9\times10^7$, $3\times10^8$, $9\times10^8$]  & fitted\\
$t_{\rm age}$ (kyr)     & age of the pulsar         & [197, 198, 199] & fitted\\
\hline
\end{tabular}
\begin{flushleft}
\footnotesize
\item Reference: (1) \cite{2002MNRAS.335..275M}
\label{tab:halo}
\end{flushleft}
\end{table*}

\begin{figure*}
  \centering
	\includegraphics[width=6.4cm]{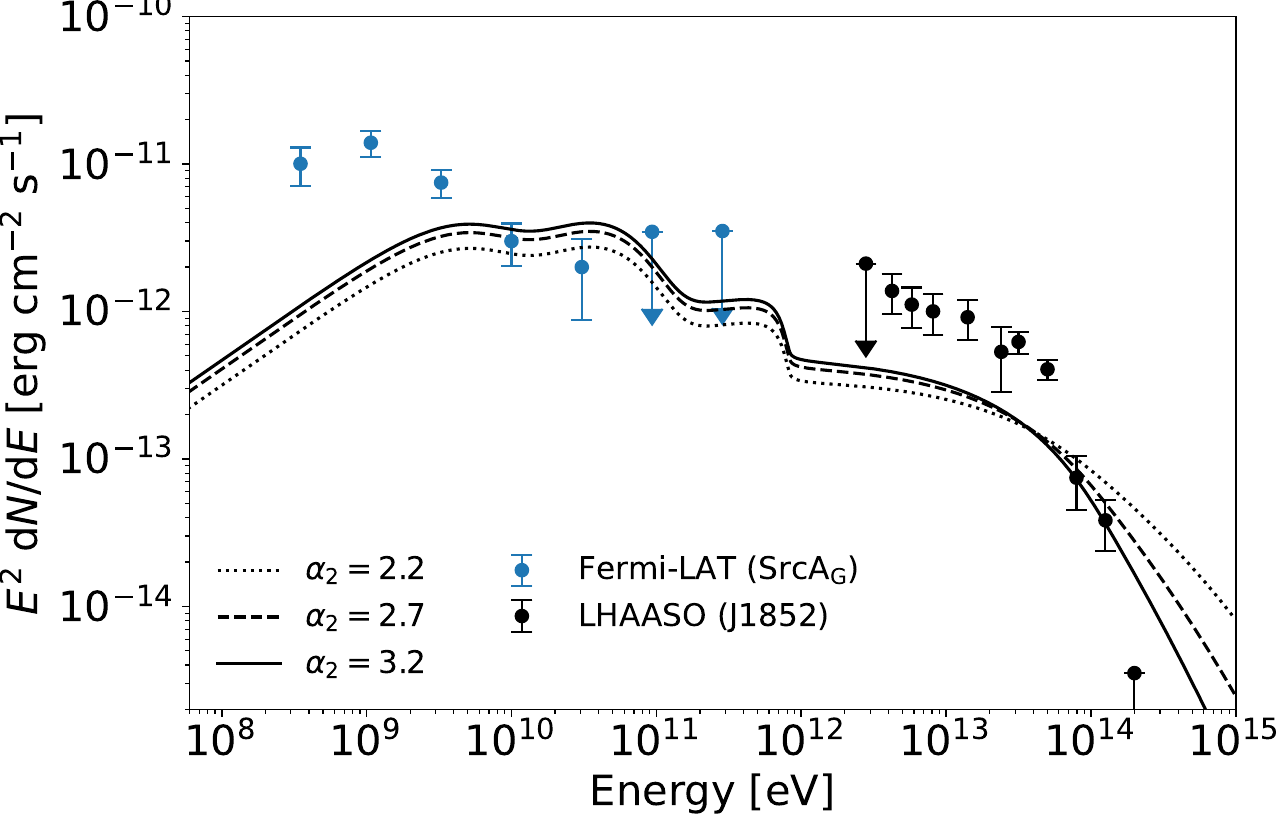}
	\includegraphics[width=5.5cm]{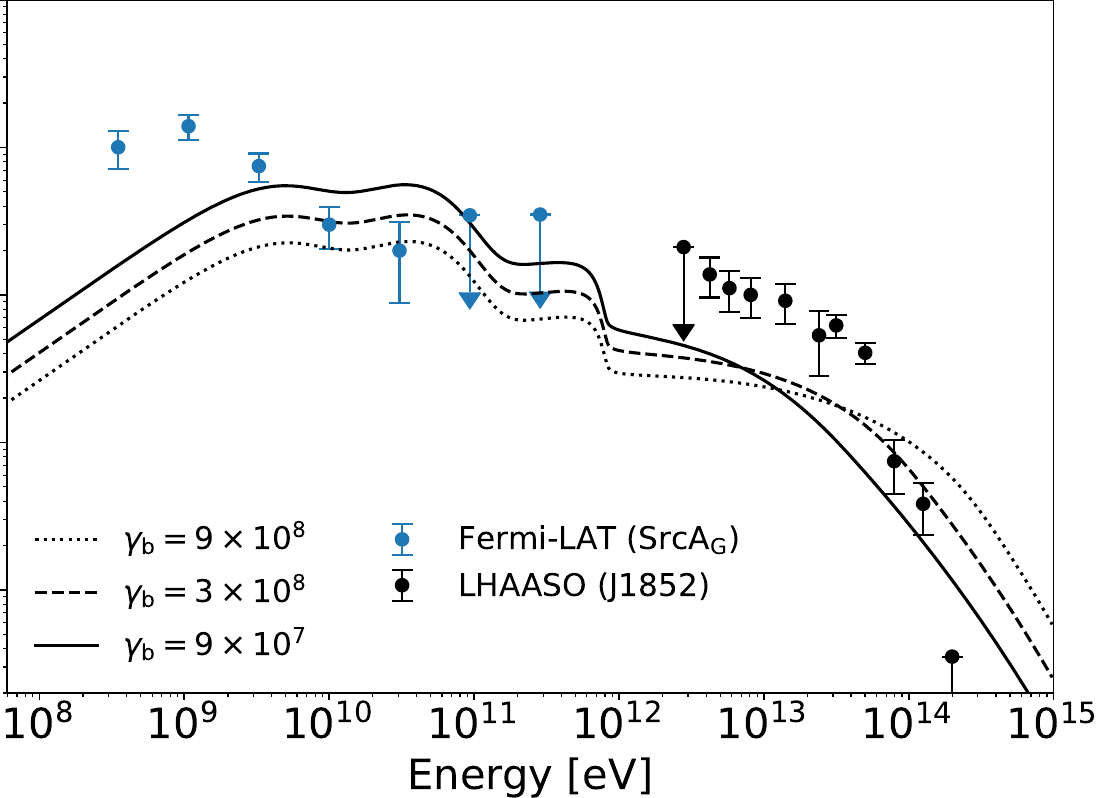}
	\includegraphics[width=5.5cm]{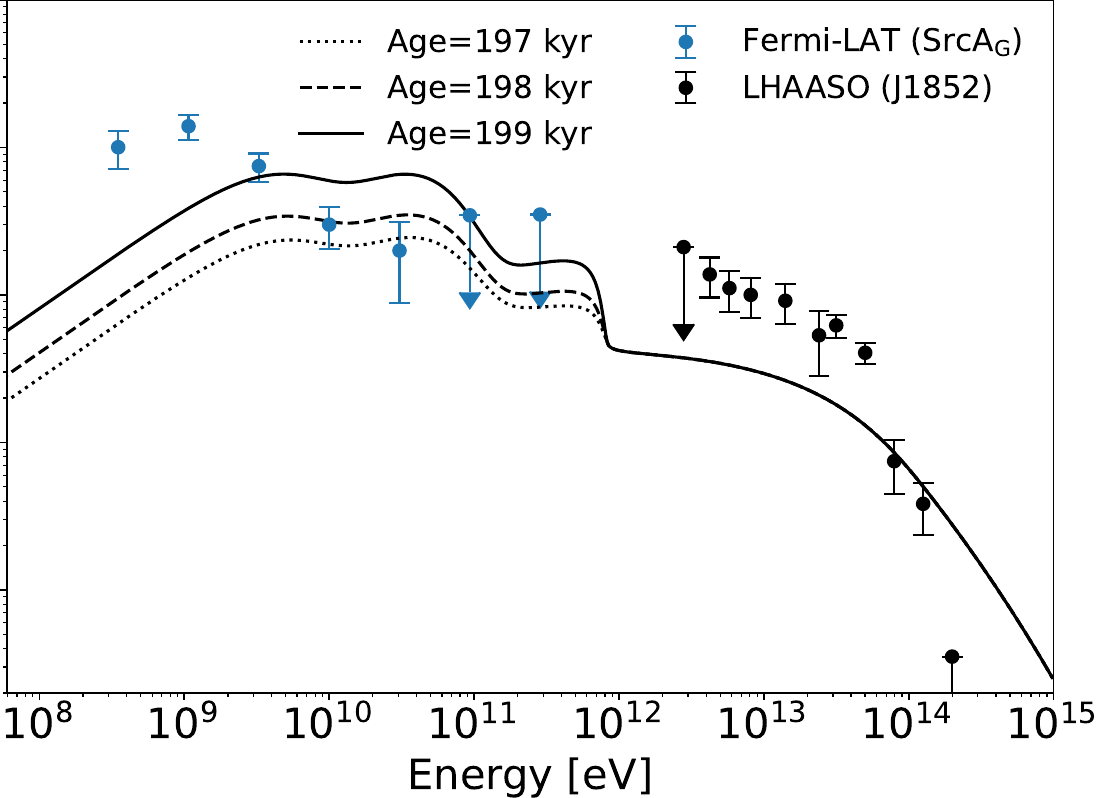}
  \caption{The flux points are same as those in \cref{fig:sed}. 
  The panels from left to right show the spectral energy distribution as functions of index $\alpha_2$, break energy $\gamma_{\rm b}$, and age $t_{\rm age}$, respectively. Left panel: the other parameters is
$\gamma_{\rm b}=3\times10^8$ and $t_{\rm age}=198$~kyr. Middle panel: $\alpha_2=2.7$ and $t_{\rm age}=198$~kyr. Right panel: $\alpha_2=2.7$ and $\gamma_{\rm b}=3\times10^8$.
  }
  \label{fig:sed_halo}
\end{figure*}

The produced distributions of gamma-ray emission by varying index, break energy, and age were shown in \cref{fig:sed_halo}.
With $\alpha_2$ getting bigger and $\gamma_{\rm b}$ getting smaller, the proportion of the electrons below break energy increases significantly.
For a given current spin-down power of the pulsar, the initial luminosity increases as the current age of the pulsar approaches the characteristic age (that is, $\tau_0$ decreases; see \cref{eq:tau0}) and further affect the current gamma-ray flux.
As can be seen in \cref{fig:sed_halo}, all of the model curves show different shapes from the observed SED and appear to peak at $\sim40$~GeV, above which the model fluxes are about an order of magnitude lower than the observed TeV flux of J1852.
Below the peak the model fluxes are also lower than the observed GeV flux of SrcA$\rm _G$.
Therefore, if it exists, the halo of this pulsar cannot account for the observed gamma-ray flux either for J1852 or SrcA$\rm _G$.

\bsp	
\label{lastpage}
\end{document}